\documentclass[11pt,a4paper]{article}
\usepackage{graphicx,amsfonts}

\newcommand{\be}{\begin{equation}}
\newcommand{\ee}{\end{equation}}
\newcommand{\bea}{\begin{eqnarray}}
\newcommand{\eea}{\end{eqnarray}}
\newcommand{\bdm}{\begin{displaymath}}
\newcommand{\edm}{\end{displaymath}}
\newcommand{\ul}{\underline}

\newcommand{\diag}{\mbox{diag}}

\newcommand{\diff}{d}
\newcommand{\Diff}{D}

\newcommand{\p}{\partial}

\newcommand{\Trace}{\mbox{Tr}}
\newcommand{\sprod}[2]{\langle #1\, , \,#2 \rangle}

\newcommand{\identy}{1 \! \! 1}

\newcommand{\const}{\mbox{const.}}

\newcommand{\bbb}{\mbox{\boldmath $b$}}
\newcommand{\bbA}{\mbox{\boldmath $A$}}
\newcommand{\bbB}{\mbox{\boldmath $B$}}
\newcommand{\bbC}{\mbox{\boldmath $C$}}
\newcommand{\bbD}{\mbox{\boldmath $D$}}
\newcommand{\bbE}{\mbox{\boldmath $E$}}
\newcommand{\bbF}{\mbox{\boldmath $F$}}

\newcommand{\bbI}{\mbox{\boldmath $I$}}
\newcommand{\bbK}{\mbox{\boldmath $K$}}
\newcommand{\bbL}{\mbox{\boldmath $L$}}
\newcommand{\bbM}{\mbox{\boldmath $M$}}
\newcommand{\bbP}{\mbox{\boldmath $P$}}
\newcommand{\bbQ}{\mbox{\boldmath $Q$}}
\newcommand{\bbR}{\mbox{\boldmath $R$}}
\newcommand{\bbS}{\mbox{\boldmath $S$}}
\newcommand{\bbT}{\mbox{\boldmath $T$}}

\newcommand{\bbX}{\mbox{\boldmath $X$}}
\newcommand{\bbY}{\mbox{\boldmath $Y$}}

\newcommand{\vep}{\ensuremath{\varepsilon}}

\newcommand{\taur}{\tau_r}

\newcommand{\gtens}{\mbox{\boldmath $g$}}

\newcommand{\gbartens}{\mbox{\boldmath $\bar{g}$}}

\newcommand{\ghat}{\hat{g}}
\newcommand{\gbar}{\bar{g}}
\newcommand{\Abar}{\bar{A}}

\newcommand{\nabhat}{\hat{\nabla}}

\newcommand{\nabbar}{\bar{\nabla}}

\newcommand{\Diffbar}{\bar{\Diff}}

\def\boxdal{\hbox{\hskip 0.5mm\hbox{\vrule width2.3mm height0.2mm
\vbox{\hrule width0.3mm height2.6mm}\hskip
-2.6mm \vbox{\hbox{\vrule width2.6mm height0.1mm}
\vskip -0.1mm\hrule width0.1mm height2.6mm}}\hskip 0.5mm}}

\begin{document}

\title{On the linear stability of solitons and
hairy black holes with a negative
cosmological constant: the even-parity sector}

\author{E. Winstanley$^{\star}$ and O. Sarbach$^{\dagger}$\\[0.2cm]
        $^{\star}$Department of Applied Mathematics,
        The University of Sheffield,\\
        Hicks Building, Hounsfield Road, Sheffield, S3 7RH, U.K.\\
        {\em {E.Winstanley@sheffield.ac.uk}}\\[0.2cm]
        $^{\dagger}$Center for Gravitational Physics and Geometry,\\
        Department of Physics, The Pennsylvania State University,\\
        104 Davey Laboratory, University Park, PA 16802.\\
        {\em {sarbach@gravity.phys.psu.edu}}}

\date{\today}

\maketitle

\begin{abstract}
Using a recently developed perturbation formalism based on
curvature quantities, we complete our
investigation of the linear stability of black holes
and solitons with Yang-Mills hair and a negative cosmological
constant.
We show that those solutions which have no linear instabilities
under odd- and even-parity spherically symmetric perturbations
remain stable under even-parity, linear, non-spherically symmetric
perturbations. Together with the result from a previous work, we
have therefore established the existence of stable hairy black holes
and solitons with anti-de Sitter asymptotic.
\end{abstract}

\section{Introduction}

Soliton and hairy black hole solutions of general relativity coupled
to non-Abelian gauge fields
have been the subject of intensive investigation for over
ten years (see \cite{review} for a comprehensive review of the
subject).
The field was sparked by the discovery of solitonic \cite{bartnik}
and coloured black hole \cite{bizon} solutions to
${\mathfrak {su}}(2)$ Einstein-Yang-Mills (EYM) theory.
Since then, many examples of both solitons and black holes
in various theories involving non-Abelian gauge fields have been
discovered, both with and without a cosmological constant.

Many of these examples have been shown to be classically unstable,
making the search for stable solitons and hairy black holes vital
in this area. An important (and surprising) discovery was that
there are solitons and black holes in ${\mathfrak {su}}(2)$ EYM
with a negative cosmological constant which are stable with
respect to linear, spherically symmetric, perturbations
\cite{W-Stable,bjork}. This result was surprising because all such
solutions in EYM with either a positive or zero cosmological
constant must be unstable \cite{brod2,brod1}. For those solutions
which are stable with respect to spherically symmetric
perturbations, it remains to be seen whether the stability is
altered if non-spherically symmetric perturbations are considered.
The perturbations fall into two sectors, depending on their
behaviour under parity transformations. In \cite{SW-Stable} we
analyzed the odd-parity sector of perturbations and showed that
those solutions which are stable under spherically symmetric
perturbations remain stable under odd-parity, linear
perturbations. The proof of stability will therefore be complete
if we can show that there are no instabilities in the even-parity
sector. This is the subject of the present article.

In common with \cite{SW-Stable}, we shall use a recently developed
perturbation formalism based on curvature quantities. The main
advantage of this formalism \cite{BHS-Letter, SHB-PRD} is that on
a static and purely magnetic background it yields a wave equation
for the linearized extrinsic curvature and electric YM field where
the spatial part of the wave operator is (formally) self-adjoint.
This offers the possibility of studying the linear stability of
non-rotating black holes and solitons by analytical means. For
example, we can use the nodal theorem \cite{AQ-Nodal} which in
fact will turn out to be an essential tool in our stability proof.
Even with these powerful tools at hand, the even-parity sector is
considerably less amenable to analysis than the odd-parity sector.
As we have shown in our previous work \cite{SW-Stable}, the
stability in the odd-parity sector is {\it topological } in the
sense that we do not need to know the exact details of the
background solutions in order to show that the system is stable.
In the even-parity sector, we will see that the detailed
structure of the metric is needed even in some simple cases.
Nonetheless, we will present an analytic proof that those solitons
and black holes having no instabilities under spherically
symmetric perturbations remain stable under even-parity linear
perturbations. In short, all soliton and black hole solutions when
the cosmological constant is sufficiently large and negative are
linearly stable.

This work is organized as follows. In section \ref{Sect-2} we
remember some important results about the hairy black hole
solutions with a negative cosmological constant, recently found in
\cite{W-Stable}, and the corresponding solitonic solutions
\cite{bjork}. In section \ref{Sect-3}, we briefly review the
curvature-based formalism of perturbation theory for a static
background. The harmonic decomposition is performed in section
\ref{Sect-4}. Special cases, such as the stability of the
Schwarzschild-anti-de Sitter and the
Reissner-Nordstr\"om-anti-de Sitter metric are discussed in
section \ref{Sect-5}. In section \ref{Sect-6}, we discuss the
general case and prove that the system is stable when $|\Lambda|$
is large enough. Given that the proof is rather involved, we first
present an outline of our argument in section \ref{Sect-6.1}
before working through each step in detail. Technical details such
as the discussion of the perturbation potential in the
Reissner-Nordstr\"om-anti-de Sitter case and the factorization of
the spatial perturbation operator in the odd-parity sector are
discussed in appendices \ref{App-A} and \ref{App-B}, respectively.

The metric signature is $(-,+,+,+)$ throughout, and we use the
standard notations $2\omega_{(ab)} = \omega_{ab} + \omega_{ba}$
and $2\omega_{[ab]} = \omega_{ab} - \omega_{ba}$ for
symmetrizing and antisymmetrizing, respectively.
Throughout the paper, greek letters denote
spacetime indices taking values
in $(0,1,2,3)$, while roman letters will denote spatial
indices taking values $(1,2,3)$.

\section{Solitons and
hairy black holes with a negative cosmological constant}
\label{Sect-2}

In this section we discuss very briefly those details of
the spherically symmetric black hole and soliton solutions of
${\mathfrak {su}}(2)$ Einstein-Yang-Mills
theory with a negative cosmological constant that we
require later in this paper.
Further details can be found in \cite{SW-Stable} and
the original papers \cite{W-Stable,bjork}.

The equilibrium metric is spherically symmetric
\bdm
ds^{2}= -N(r)S^{2}(r)\, dt^{2}+N^{-1}(r)\, dr^{2} +r^{2}
\left( d\theta ^{2} +\sin ^{2} \theta \, d\phi ^{2} \right) ,
\edm
and the gauge field potential has the spherically symmetric
form
\bdm
A=(1-w(r)) \left[ -\tau _{\phi } \diff \theta + \tau _{\theta }
\sin \theta \, \diff\phi  \right] .
\edm
Here the ${\mathfrak {su}}(2)$ generators $\tau _{r,\theta ,\phi }$
are given in terms of the usual Pauli matrices $\sigma _{i}$
by $\tau _{r}={\ul {e}}_{r}\cdot {\ul {\sigma }}/2i$, etc.
Writing $N(r)=1-2m(r)/r-\Lambda r^{2}/3$, where $\Lambda $
is the (negative) cosmological constant,
the field equations take the form:
\bea
m_{,r}& = & G\left[ Nw_{,r}^{2} +\frac {1}{2r^{2}}(w^{2}-1)^{2}
\right] ,
\nonumber \\
\frac {S_{,r}}{S} & = & \frac {2G w_{,r}^{2}}{r} ,
\nonumber \\
0 & = & Nr^{2}w_{,rr}+\left(
2m -\frac {2\Lambda r^{3}}{3} -
G \frac {(w^{2}-1)^{2}}{r} \right) w_{,r} +(1-w^{2})w,
\label{eq-equil}
\eea
where $G$ is Newton's constant, and we have set the gauge
coupling constant equal to $\sqrt{4\pi}$ for convenience.
Here, and in the rest of the paper, we use a comma in a subscript
to denote partial differentiation, i.e. $w_{,r}=\p _{r}w$.

For solutions which approach anti-de Sitter (adS)
space at infinity, the asymptotic behaviour of the
field functions is:
\bea
m(r) & = & M + \frac {M_{1}}{r}+O(r^{-2}) ,
\nonumber \\
w(r) & = & w_{\infty }+\frac {w_{1}}{r} +O(r^{-2}) ,
\nonumber \\
S(r) & = & 1 + O(r^{-4}).
\nonumber
\eea
Due to these boundary conditions, in general the
solutions will be globally magnetically charged.

For black hole solutions having a regular event horizon at
$r=r_{h}$, all the field variables have regular Taylor
expansions near the event horizon, for example
\bdm
w(r)=w(r_{h})+w_{,r}(r_{h}) (r-r_{h})+O(r-r_{h})^{2} .
\edm
However, there are just two independent parameters, $S(r_{h})$
and $w(r_{h})$ since $N=0$ at the event horizon, which gives
\bdm
m(r_{h})= \frac {r_{h}}{2} -\frac {\Lambda r_{h}^{3}}{6}\, .
\edm
In order for the event horizon to be regular, we shall also require
that $N_{,r}(r_h)>0$, which implies that
\bdm
F_h \equiv 1 - \Lambda r_h^2 - G\frac{(w(r_h) - 1)^2}{r_h^2} > 0.
\edm
From (\ref{eq-equil}), one has
\bdm
w_{,r}(r_{h})=\frac {(w(r_{h})^{2}-1)w(r_{h})}{r_{h} F_h}\, .
\edm
There are also globally regular (solitonic) solutions,
for which the behaviour near the origin is:
\bea
m(r) & = & 2G b^{2} r^{3}+O(r^{4}) ,
\nonumber \\
w(r) & = & 1-br^{2}+O(r^{3}) ,
\nonumber \\
S(r) & = & S(0) \left[ 1+4G b^{2} r^{2} +O(r^{3}) \right].
\nonumber
\eea
Here the independent parameters are $b$ and $S(0)$.

The simplest solutions to the field equations (\ref{eq-equil})
are the Schwarzschild-adS solution,
\bdm
w=\pm 1, \;\;\; S=1, \;\;\; m=\const
\edm
and the Reissner-Nordstr\"om-adS (RN-adS) solution
\bdm
w=0, \;\;\; S=1, \;\;\; m=\const - \frac{G}{2r}\, .
\edm
In both cases, the YM field is effectively Abelian.
It is however interesting to study the stability of
these solutions with respect to non-Abelian perturbations
(see section \ref{Sect-5}).

The solutions of greatest interest in this article
are those effectively non-Abelian solutions,
for which the gauge function $w$ has
no zeros, since these solutions were shown in
\cite{W-Stable} to be linearly stable to both
even- and odd-parity spherical perturbations.
In \cite{W-Stable} it is proved that for any value
of the gauge field at the event horizon, $w(r_{h})\neq 0$,
for all sufficiently large $|\Lambda |$ there is a black hole
solution in which $w$ has no zeros.
Similar behaviour is found numerically for the solitonic
solutions \cite{bjork}.
For spherical perturbations of these equilibrium configurations,
it is proved analytically that all the solutions in which $w$ has
no zeros are stable in the odd-parity sector \cite{W-Stable}.
The even-parity sector is more complicated, but stability
can be proven for sufficiently large $|\Lambda |$.

In addition to the boundary conditions on the field variables
at the origin (or event horizon, as applicable) and infinity,
we shall in section \ref{Sect-6} make extensive use of
the fact that the equilibrium solutions (both solitonic and
black hole) are analytic in both $r$ and the parameter $\Lambda $.
This is proved in \cite{W-Stable} for black hole solutions,
a proof which is readily extended to cover the solitonic case
(see, for example \cite{breit} for the asymptotically flat situation).
We shall also apply the result, proven in \cite{W-Stable},
that
\bdm
w_{,r}(r) \sim o(|\Lambda |^{-\frac {1}{2}})
\qquad
{\mbox {as $|\Lambda |\rightarrow \infty $.}}
\edm

Dyonic solutions with a non-vanishing electric field also
exist in this model with $\Lambda <0$ \cite{bjork}.
However, in common with our analysis of the odd-parity perturbations
\cite{SW-Stable}, our work here applies only to the purely
magnetic equilibrium solutions.
In addition, recently hairy black holes
with non-spherical
event horizon topology have been found in this model
\cite{bij}.
Here we consider only black holes whose event horizon
has spherical topology, although we conjecture that
all black holes (whatever the topology of their event horizon)
will be stable if $|\Lambda |$ is sufficiently large.

\section{The pulsation equations}
\label{Sect-3}

Generalizing previous results \cite{BHS-Letter, SHB-PRD}, we have
shown in \cite{SW-Stable} that linear fluctuations on a static
and purely magnetic background are governed by a symmetric wave
equation for the linearized extrinsic curvature and electric YM field.
Performing an ADM decomposition of the metric and the gauge
potential,
\bea
\gtens &=& -\alpha^2 \diff t^2 + \gbar_{ij} (\diff x^i +
\beta^i \diff t)(\diff x^j + \beta^j \diff t),
\nonumber\\
A &=& -\Phi\, \alpha\diff t + \Abar_i (\diff x^i + \beta^i \diff t).
\nonumber
\eea
(where $\Phi$ and $\Abar_i$ are both Lie algebra valued),
these quantities are given by
\bdm
L_{ij} = \frac{1}{2}\delta \left( \p_t\gbar_{ij} - L_\beta\gbar_{ij} \right)
\edm
and
\bdm
{\cal E}_i = -\delta\left( \p_t\Abar_i + \Diffbar_i(\alpha\Phi) \right).
\edm
(We use the same notation as in \cite{SW-Stable}: In particular,
a bar refers to the 3-metric $\gbar_{ij}$ or to the magnetic potential $\Abar_i\,$.)
On the background, static coordinates are chosen such that
the shift $\beta^i$ and the time derivative of the 3-metric,
$\p_t\gbar_{ij}$ vanish. Similarly, the gauge is chosen such that
the electric potential $\Phi$ and the time-derivative of the
magnetic potential, $\p_t\Abar_i$ vanish. As a consequence,
$L_{ij}$ and ${\cal E}_i$ are vector-invariant, that is, invariant
with respect to both infinitesimal coordinate transformations
within the ADM slices and infinitesimal gauge transformations of
the gauge potential. More precisely, with respect to a coordinate
transformation $\delta x^\mu \mapsto \delta x^\mu + X^\mu$, we have
\bea
L_{ij} &\mapsto& L_{ij} + \nabbar_{(i} \left( \alpha^2 \nabbar_{j)} f \right),
\label{Eq-GravGaugeFreedom}\\
{\cal E}_i &\mapsto& {\cal E}_i + \alpha^2\bar{F}_{ij} \nabbar^j f\, ,
\label{Eq-YMGaugeFreedom}
\eea
where $f = X^t$. The fact that the spatial components, $X^i$, do
not appear in (\ref{Eq-GravGaugeFreedom}) and (\ref{Eq-YMGaugeFreedom})
justifies the name ``vector-invariant''.

The pulsation equations can be obtained from the following energy functional:
\bdm
E = E_{grav} + E_{YM} + E_{int}\, ,
\edm
where
\bea
E_{grav} &=& \frac{1}{2} \int\left(
     \frac{\bar{G}^{ijkl}}{\alpha^2} \dot{L}_{ij}\dot{L}_{kl}
   + \bar{G}^{ijkl}(\nabbar^s L_{ij})(\nabbar_s L_{kl})
   + 2L^{ij} \bar{R}^k_{\; i} L_{jk}
   - 2L^{ij} \bar{R}_{kilj} L^{kl} \right. \nonumber\\
  && \left. +\, \frac{4}{\alpha} L^{ij}\nabbar_i \left( L_{jk}\nabbar^k\alpha \right)
   -\frac{4}{\alpha} L^{ij}(\nabbar_i\alpha) \nabbar^k L_{kj}
   - 2L^{ij}\nabbar^k \left( \frac{\nabbar_i\alpha}{\alpha} \right) L_{jk}
   \right. \nonumber\\
  && \left. -\, \frac{4}{\alpha^2} L \nabbar^i(\alpha\nabbar^j\alpha) L_{ij}
   + \frac{2}{\alpha^2}\, (\nabbar^k\alpha)(\nabbar_k\alpha) L^2
   - \frac{2}{\alpha^2}\bar{G}^{ijkl} L_{ij} \nabbar_k\alpha^2\nabbar_l \tilde{A} \right.
   \nonumber\\
  && \left. -\, 2\Lambda L^{ij} L_{ij}
   + 4G\Trace\left\{ L^{ij}\bar{F}^k_{\;\, i} \bar{F}^l_{\; j} L_{kl}
   + \frac{1}{4} \bar{F}_{kl} \bar{F}^{kl} L_{ij} L^{ij}
   \right\}\right) \alpha\sqrt{\gbar}\,\diff\,x^3, \nonumber\\
E_{YM} &=& G \int\Trace\left\{ \frac{1}{\alpha^2} \dot{\cal E}^i\dot{\cal E}_i
 + 2\Diffbar^{[i}{\cal E}^{j]}\cdot \Diffbar_{[i}{\cal E}_{j]}
 + \alpha^2 \left[ \Diffbar^j\left(
     \frac{{\cal E}_j}{\alpha} \right) \right]^2 \right. \nonumber\\
&& \left. \qquad\qquad +\, \bar{F}^{ij}[{\cal E}_i, {\cal E}_j]
 + 4G\Trace( \bar{F}^l_{\; k} {\cal E}_l )
 \bar{F}^{ik}{\cal E}_i \right\} \alpha\sqrt{\gbar}\,\diff\,x^3, \nonumber\\
E_{int} &=& -2G\int\Trace\left\{
    L^{ij}\Diffbar_k \left( \alpha\bar{F}_i^{\; k} {\cal E}_j \right)
 - \alpha \bar{F}^{jk} {\cal E}^i \nabbar_k L_{ij}
 + \frac{2}{\alpha} L^{ij} {\cal E}_k
   \Diffbar_i \left( \alpha^2\bar{F}_j^{\; k} \right) \right. \nonumber\\
&& \left. \qquad\qquad -\, 2\bar{F}^{kl}(\nabbar_k\alpha) L {\cal E}_l
 + \alpha\bar{F}_{ij}{\cal E}^i \nabbar^j \tilde{A} \right\} \sqrt{\gbar}\,\diff\,x^3.
\nonumber
\eea
Here, $\tilde{A}$ is defined by
\be
\tilde{A} = \frac{\delta\dot{\alpha} - \delta\beta_j\nabbar^j\alpha}{\alpha} - L,
\label{Eq-LinDensLapse}
\ee
which is the difference between a vector-invariant combination of
the time-derivative of the lapse and the trace, $L = \gbar^{ij} L_{ij}\,$,
of the extrinsic curvature.
By absorbing $L$ into $\tilde{A}$, one makes sure that the
resulting dynamical equations are hyperbolic. In fact, it can be
checked that (\ref{Eq-LinDensLapse}) corresponds to a linearized
version of a densitized lapse, which is widely used in recent
hyperbolic formulations of Einstein's equations
(see Ref. \cite{Hyperbolic} for example).

In terms of the variable $\tilde{A}$, the harmonic gauge reads
$\tilde{A} = 0$. In this case, one has a set of equations with
symmetric potential, but the kinetic energy is not positive
\cite{BHS-Letter}. This is the reason why the De Witt metric
\bdm
\bar{G}^{ijkl} = \gbar^{i(k} \gbar^{l)j} -  \gbar^{ij}\gbar^{kl}
\edm
appears in $E_{grav}\,$.

As we have discussed in \cite{SHB-PRD}, one can also adopt the
maximal gauge, $L=0$, in which case the equations are both
hyperbolic and symmetric, but where perturbations of the lapse
still appear in $\tilde{A}$. In this case, one recovers the
energy functional given in Appendix A of Ref. \cite{SW-Stable}.
At this point we recall that for perturbations with odd parity,
$L = 0$ and $\tilde{A} = 0$ anyway since they represent scalar quantities.
In the even-parity sector, it will turn out to be useful not to choose
a particular gauge {\it {a priori}} but to derive the pulsation equations in
a general gauge and to choose an appropriate gauge later.

The dynamical equations resulting from the variation of $E$ are
subject to the linearized momentum and Gauss constraint equations,
\bea
0 = {\cal C}_i & \equiv & \alpha\bar{G}_i^{\; jkl} \nabbar_j\left( \frac{L_{kl}}{\alpha} \right)
    - 2G\Trace\left( \bar{F}^j_{\; i} {\cal E}_j \right), \nonumber\\
0 = {\cal G} & \equiv & \alpha\Diffbar^j\left( \frac{{\cal
      E}_j}{\alpha} \right).
\nonumber
\eea
Additional constraints involving also perturbations of the metric
and the gauge potential themselves are the Hamiltonian constraint
and all evolution equations,
which we had differentiated in time in order to construct the
wave operator.

Finally, we will make use of the following fact \cite{SHB-PRD} later
in our stability analysis: In any gauge, the terms involving $\tilde{A}$
in the energy functional do not contribute provided that $L_{ij}$
and ${\cal E}_i$ satisfy the momentum constraint ${\cal C}_ i = 0$.
Indeed, using partial integration (and homogeneous Dirichlet boundary
conditions for the perturbations), the terms involving $\tilde{A}$ are
\bea
&& \int\left( -\frac{1}{\alpha}\bar{G}^{ijkl} L_{ij} \nabbar_k\alpha^2\nabbar_l \tilde{A}
  -2G\alpha\Trace\left\{ \bar{F}_{ij}{\cal E}^i \nabbar^j \tilde{A} \right\} \right) \sqrt{\gbar}\,\diff\,x^3
\nonumber\\
&& = \int\alpha\left( \alpha\nabbar_k\left\{ \bar{G}^{ijkl}\frac{L_{ij}}{\alpha} \right\}
   - 2G\Trace\left\{ \bar{F}^{kl}{\cal E}_k \right\} \right) \nabbar_l \tilde{A}
\sqrt{\gbar}\,\diff\,x^3 ,
\nonumber
\eea
which vanishes if ${\cal C}^l = 0$.

\section{Even-parity fluctuations}
\label{Sect-4}

Here, we specialize the pulsation equations given in the last
section to a spherically symmetric background with gauge group
${\mathfrak {su}}(2)$. In this case, it is convenient to expand
the linearized extrinsic curvature and electric YM field in terms
of spherical tensor harmonics since then, perturbations belonging
to different choices of the angular momentum numbers $\ell$ and
$m$ decouple. Furthermore, the tensor harmonics can be divided
into parities.
Odd-parity perturbations were discussed in \cite{SW-Stable}.
In this work, we consider the even-parity sector.

\subsection{The vacuum case}

We start by computing $E_{grav}\,$.
For technical reasons, it is convenient to parametrize the background
$3$-metric according to
\bdm
\gbartens = \diff x^2 + r(x)^2 \diff\Omega^2,
\edm
where $x$ is a radial coordinate, and where $r$ and
the lapse $\alpha$ depend on $x$ only. In terms of the functions $N$
and $S$ defined in section \ref{Sect-2}, we have
$\alpha^2 = N S^2$ and $N = (\p_x r)^2$.

Einstein's background vacuum equations are obtained from
\bea
\frac{(r^2\alpha')'}{r^2\alpha} = &R_{00}& = \frac{2G}{r^2} \left( w'^2 + \frac{(w^2-1)^2}{2r^2} \right) - \Lambda,
\nonumber\\
-2\frac{r''}{r} - \frac{\alpha''}{\alpha} = &R_{xx}& = \frac{2G}{r^2} \left( w'^2 - \frac{(w^2-1)^2}{2r^2} \right) + \Lambda,
\nonumber \\
2\left( 1 - r'^2 - r r'' - r r'\frac{\alpha'}{\alpha} \right) = &R^A_{\; A}& = \frac{2G}{r^2} \frac{(w^2-1)^2}{r^2} + 2\Lambda.
\label{Eq-ADMBackSph}
\eea
Here and in the following, capital indices refer to coordinates on the
$2$-sphere, and a prime denotes differentiation with respect to $x$.

In the even-parity sector, we expand $L_{ij}$ according to
\bea
&& L_{xx} = \left( \tilde{p} + \frac{1}{3}\tilde{t} \right) e^{(1)}, \nonumber\\
&& L_{xB} = \tilde{q}\,e^{(2)}_B\, ,
\nonumber
\\
&& L_{AB} = \left( -\frac{1}{2}\tilde{p} + \frac{1}{3}\tilde{t} \right) e^{(3)}_{AB}
          + \tilde{g}\, e^{(4)}_{AB} \nonumber,
\eea
where
\bea
&& e^{(1)} = \frac{Y}{r}\, ,\;\;\;
   e^{(2)}_A = \nabhat_A Y, \nonumber\\
&& e^{(3)}_{AB} = r\ghat_{AB} Y, \;\;\;
   e^{(4)}_{AB} = r \left( \nabhat_A\nabhat_B Y + \frac{1}{2}\ell(\ell+1)\ghat_{AB} Y \right),
\nonumber
\eea
form a basis of even-parity tensor harmonics, which are orthogonal
with respect to the inner product induced by $\gbartens$.
($Y \equiv Y^{\ell m}$ denote the standard spherical
harmonics. The indices $\ell m$ are suppressed in what follows.)
Also, we have chosen the parametrization such that $\tilde{t}$ and $\tilde{p}$
correspond to the trace and the radial trace-less part, respectively,
of $L_{ij}$.
After the rescaling
\bdm
\tilde{t} = \sqrt{\frac{3}{2}}\, t, \;\;\;
\tilde{p} = \sqrt{\frac{2}{3}}\, p, \;\;\;
\tilde{q} = \frac{1}{\sqrt{2\mu^2}}\, q, \;\;\;
\tilde{g} = \sqrt{\frac{2}{\mu^2\lambda}}\, g,
\edm
with $\mu^2 = \ell(\ell+1)$, $\lambda = \mu^2 - 2$, the expansion
is normalized such that the De Witt metric becomes
\bdm
\int L_{ij}\bar{G}^{ijkl} L_{kl} \sqrt{\gbar}\,\diff\,x^3
 = \int\limits (-t^2 + p^2 + q^2 + g^2) \diff x.
\edm
The signs reflect the signature of the De Witt metric.

Inserting the expressions above, using the background quantities
\bea
& \bar{R}^x_{\; AxB} = -rr'' \ghat_{AB},
& \bar{R}^D_{\; CAB} = 2(1 - r'^2) \delta^D_{\; [A} \ghat_{B]C},
\nonumber\\
& \bar{R}_{xx} = -2\frac{r''}{r}\, ,
& \bar{R}_{AB} = \left( 1 - r'^2 - rr'' \right)\ghat_{AB},
\nonumber\\
& \bar{F}_{xB} = -w' \hat{\vep}^A_{\; B} \tau_A\, ,
& \bar{F}_{AB} = (w^2 - 1)\taur\, \hat{\vep}_{AB}\, ,
\nonumber
\eea
and integrating over the spherical variables,
the gravitational energy functional becomes, after some calculations,
\bdm
E_{grav} = \frac{1}{2} \int
   \left( \sprod{\dot{V}}{\bbT\dot{V}} + \sprod{\p_\rho V}{\bbT\p_\rho V}
 + \sprod{V}{\bbS_{grav} V} - 2\sprod{V}{\bbT\bbb(\tilde{a})_{grav}} \right) \diff\rho,
\edm
where $V\equiv (t,p,q,g)^T$. The matrix $\bbT$ is given by
\bdm
\bbT = \diag(-1,1,1,1)
\edm
and the symmetric matrix $\bbS_{grav}$ by
\bdm
\bbS_{grav} = \left( \begin{array}{cccc}
  S_{tt} & S_{t1} & 0 & 0 \\
  S_{t1} & S_{11} & \sqrt{12}\mu\gamma_{,\rho }& 0 \\
  0 & \sqrt{12}\mu\gamma_{,\rho } & S_{22}
  & 2\sqrt{\lambda}\gamma_{,\rho }\\
  0 & 0 & 2\sqrt{\lambda}\gamma_{,\rho }& S_{33}
  \end{array} \right),
\edm
with
\bea
S_{t1} &=& -\frac{4r}{3}(\gamma\alpha')' + \frac{4G}{3}\gamma^2\left[ w'^2 - \frac{(w^2-1)^2}{r^2} \right],
\nonumber\\
S_{tt} &=& -\gamma^2\left[ \mu^2 + rr'' + \frac{13}{3} rr'\frac{\alpha'}{\alpha}
        + \frac{5}{3}\frac{r^2}{\alpha}\,\alpha''
        - \frac{4}{3}\frac{r^2}{\alpha^2}\, \alpha'^2 + \Lambda r^2
       \right] \nonumber\\
       & & +\frac{14G}{3}\gamma^2\left[ w'^2 + \frac{(w^2-1)^2}{2r^2} \right], \nonumber\\
S_{11} &=& \gamma^2\left[ \mu^2 + 6r'^2 - 5rr'' - \frac{31}{3} rr'\frac{\alpha'}{\alpha}
        + \frac{4}{3}\frac{r^2}{\alpha^2}(\alpha\alpha')' - 2\Lambda r^2 \right] \nonumber\\
       & & - \frac{4G}{3}\gamma^2\left[ w'^2 - 5\frac{(w^2-1)^2}{2r^2} \right], \nonumber\\
S_{22} &=& \gamma^2\left[ \mu^2 + 4r'^2 - 4rr'' - 8 rr'\frac{\alpha'}{\alpha}
        + \frac{r^2}{\alpha^2}(\alpha\alpha')' - 2\Lambda r^2  + 2G\frac{(w^2-1)^2}{r^2} \right],
       \nonumber\\
S_{33} &=& \gamma^2\left[ \mu^2 - 2r'^2 - rr'' - rr'\frac{\alpha'}{\alpha} - 2\Lambda r^2
            + 4Gw'^2 - 2G\frac{(w^2-1)^2}{r^2} \right]. \nonumber
\eea
Here, we have also introduced $\gamma = \alpha/r$ and
$\p_\rho = \alpha\p_x\,$.
The inhomogeneous term $\bbb(\tilde{a})_{grav}\,$, where $\tilde{a} = \tilde{a}(x)$
is the scalar amplitude parametrizing $\tilde{A}$,
\bdm
\tilde{A} = \tilde{a} Y,
\edm
is given by $\bbb(\tilde{a})_{grav} = (b_t, b_1, b_2, b_3)(\tilde{a})$, where
\bea
b_t(\tilde{a}) &=& \sqrt{\frac{2}{3}}\left[ r(\alpha^2\tilde{a}')' + 2\alpha^2 r' \tilde{a}' -
             \mu^2\frac{\alpha^2}{r}\,\tilde{a} \right], \nonumber\\
b_1(\tilde{a}) &=& \sqrt{\frac{2}{3}}\left[ r (\alpha^2\tilde{a}')' - \alpha^2r'\tilde{a}'
          + \frac{\mu^2}{2}\frac{\alpha^2}{r}\,\tilde{a} \right], \nonumber\\
b_2(\tilde{a}) &=& \sqrt{2\mu^2}\, \alpha r \left( \frac{\alpha}{r}\,\tilde{a} \right)', \nonumber\\
b_3(\tilde{a}) &=& \sqrt{\frac{\mu^2\lambda}{2}}\, \frac{\alpha^2}{r}\,\tilde{a}. \nonumber
\eea
The resulting perturbation equations are
\be
\left( \bbT \p_t^{\, 2} - \bbT \p_\rho^{\, 2} + \bbS_{grav} \right) V - \bbT\bbb(\tilde{a})_{grav} = 0.
\label{Eq-WaveEqVacSphEven}
\ee
The linearized momentum constraint yields the
following two equations:
\bea
0 = \alpha{\cal C}^{vac}_x
 &=& \sqrt{\frac{2}{3}}\left[ \frac{\alpha}{r^2}\p_{\rho}\left( \frac{r^2 p}{\alpha} \right)
  - \alpha r\p_{\rho} \left( \frac{t}{\alpha r} \right)
  - \frac{\sqrt{3}\mu}{2}\gamma\, q \right] \frac{Y}{r}\, ,
\nonumber \\
0 = \alpha {\cal C}^{vac}_B
 &=& \frac{1}{\sqrt{2\mu^2}}\left[ \frac{\alpha}{r^2}\p_{\rho}\left( \frac{r^2 q}{\alpha} \right)
  - \frac{2\sqrt{3}\mu}{3}\gamma\, t - \frac{\sqrt{3}\mu}{3}\gamma\, p
  - \sqrt{\lambda}\gamma\, g \right] \nabhat_B Y. \nonumber
\eea

Note that the spatial operator which acts on $V$ is symmetric (with
respect to the standard $L^2$ scalar product).
However, compared to the odd-parity case, the following problems
arise here:
First, the kinetic energy is not positive. In order to fix this,
one could multiply equation (\ref{Eq-WaveEqVacSphEven}) from the
left with $\bbT$. However, one would then lose the symmetry of
the spatial operator.
Next, the (densitized) lapse still appears into the equations,
and one has no evolution equation for $\tilde{a}$. On the other hand,
the amplitudes are not fully coordinate-invariant but are subject
to the gauge transformations (\ref{Eq-GravGaugeFreedom}).
As a consequence, we have
\be
V \mapsto V + \bbb(\xi)_{grav}
\label{Eq-VacGaugeFreedom}
\ee
where $\xi(x)$ parameterizes $f$ according to $f = \xi\, Y$.
Since the same operator-valued vector $\bbb(.)_{grav}$ appears
in both the coordinate transformation and the terms involving
$\tilde{a}$ in the dynamical equations, it is clear that any
gauge-invariant combination of the equations for $t$, $p$, $q$
and $g$ annihilates the terms which depend on $\tilde{a}$.

Using the background equations (\ref{Eq-ADMBackSph}), we can
re-express $rr''$, $r'\alpha'$ and $\alpha''$ in terms of
$N = r'^2$ and matter terms:
\bea
rr'' &=& \frac{1}{2}(1 - N) - \frac{\Lambda}{2} r^2
      - G w'^2 - \frac{G}{2} \frac{(w^2-1)^2}{r^2}\, , \nonumber\\
rr'\frac{\alpha'}{\alpha} &=& \frac{1}{2}(1 - N) - \frac{\Lambda}{2} r^2
      + G w'^2 - \frac{G}{2} \frac{(w^2-1)^2}{r^2}\, ,
\nonumber
\\
r^2\frac{\alpha''}{\alpha} &=& -(1 - N) + 2G\frac{(w^2-1)^2}{r^2}\, .
\nonumber
\eea
As a result, one can rewrite the coefficients of $\bbS_{grav}$ as
\bea
S_{t1} &=& -\frac{4}{3} \frac{\alpha_{,\rho\rho}}{\alpha} +
           \frac{2\gamma^2}{3} \left[ 1 - N - \Lambda r^2 + 4G w'^2 - 3G\frac{(w^2-1)^2}{r^2}
           \right], \nonumber\\
S_{tt} &=& -\gamma^2\left[ \lambda + 1 + N - \Lambda r^2 - 4G w'^2 + 3G\frac{(w^2-1)^2}{r^2}
           \right] - S_{t1}\, , \nonumber\\
S_{11} &=& \gamma^2\left[ \lambda - 5 + 13N + 5\Lambda r^2
        - 4Gw'^2 + 9G\frac{(w^2-1)^2}{r^2} \right] - S_{t1}\, , \nonumber\\
S_{22} &=& \frac{r^2}{\alpha} \left( \frac{\alpha}{r^2} \right)_{,\rho\rho}
        + \gamma^2\left[ \lambda + 4G\frac{(w^2-1)^2}{r^2} \right], \nonumber\\
S_{33} &=& \frac{r_{,\rho\rho}}{r}
        + \gamma^2\left[ \lambda + 4Gw'^2 \right].
\label{Eq-Sgrav}
\eea

\subsection{The pure YM case}

Next, we compute $E_{YM}$.
The background gauge potential has the components
\bdm
\bar{A}_x = 0, \;\;\;
\bar{A}_B = (1 - w) \hat{\vep}^C_{\; B}\tau_C\, ,
\edm
where $w$ depends on $x$ only.
The background YM equation reads
\bdm
\frac{1}{\alpha} \left( \alpha w' \right)' = w\frac{w^2-1}{r^2}\, .
\edm
The electric YM field is expanded into ${\mathfrak {su}}(2)$-valued
spherical harmonic one-forms with even parity (these one-forms are
explained in Appendix D of Ref. \cite{SHB-Odd}):
\bea
{\cal E}_x &=& \frac{\tilde{b}}{r}\, X_3\, ,\nonumber\\
{\cal E}_B &=& \hat{\vep}^A_{\; B} \left[ \tilde{c}\, \taur\nabhat_A Y + \tilde{d}\, Y \tau_A +
\tilde{e}\left( \nabhat_A X_2 + \frac{1}{2}\mu^2 Y \tau_A\right) \right],
\nonumber
\eea
where in terms of the Pauli matrices $\ul{\sigma} = (\sigma^i)$,
$\taur = \ul{e}_r\cdot\ul{\sigma}/(2i)$,
$\tau_A = \ul{e}_A\cdot\ul{\sigma}/(2i)$.
Here $X_2$ and $X_3$ are the ${\mathfrak {su}}(2)$-valued harmonics
\bdm
X_2 = \ghat^{AB} \nabhat_A Y \tau_B\, , \;\;\;
X_3 = \hat{\vep}^{AB} \nabhat_A Y \tau_B\, .
\edm
After the rescaling
\bdm
\tilde{b} = \frac{b}{\sqrt{2G\mu^2}}\, , \;\;\;
\tilde{c} = \frac{c}{\sqrt{2G\mu^2}}\, , \;\;\;
\tilde{d} = \frac{d}{2\sqrt{G}}\, , \;\;\;
\tilde{e} = \frac{e}{\sqrt{G\mu^2\lambda}}\, ,
\edm
the amplitudes are normalized such that
\bdm
2G\int \Trace\left( \gbar^{ij}{\cal E}_i {\cal E}_j \right) \sqrt{\gbar}\,\diff\,x^3
   = \int (b^2 + c^2 + d^2 + e^2) \diff x.
\edm
The YM energy functional becomes
\bdm
E_{YM} = \frac{1}{2} \int
  \left( \sprod{\dot{W}}{\dot{W}} + \sprod{\p_\rho W}{\p_\rho W}
    + \sprod{W}{\bbS_{YM} W} \right) \diff\rho,
\edm
where $W\equiv (b,c,d,e)^T$ and
\be
\frac{\bbS_{YM}}{\gamma^2} = \left( \begin{array}{cccc}
  \frac{\gamma_{,\rho\rho}}{\gamma^3} + \lambda + f + \frac{u^2}{\gamma^2} & sym. & sym. & sym. \\
  -\frac{2}{\gamma^2}(\gamma w)_{,\rho }- \frac{u v}{\gamma} & \lambda + 2f + v^2 & sym. & sym. \\
  \sqrt{2}\mu\frac{\gamma_{,\rho}}{\gamma^2} & \sqrt{2}\mu w & \lambda - 2 + 3f + \frac{2 u^2}{\gamma^2} & sym. \\
  \sqrt{2\lambda}\frac{\gamma_{,\rho}}{\gamma^2} & -\sqrt{2\lambda} w & 0 & \mu^2 - f
  \end{array} \right),
\label{Eq-S_YM}
\ee
with
\be
f \equiv w^2 + 1, \;\;\;
u \equiv 2\sqrt{G}\,\frac{w_{,\rho}}{r}, \;\;\;
v \equiv 2\sqrt{G}\,\frac{w^2-1}{r}\, .
\label{Eq-udef}
\ee
(Note that the $u$ here differs from the $u$ in our previous article
\cite{SW-Stable} by a factor of $2\sqrt{G}$.)
The linearized Gauss constraint gives
\bdm
0 = \sqrt{2G}\mu\alpha r {\cal G}
  = \left\{ \gamma\p_\rho\left( \frac{b}{\gamma} \right) + \gamma w c
- \frac{\mu}{\sqrt{2}}\gamma\, d - \sqrt{\frac{\lambda}{2}}\,\gamma\, e \right\} X_3\, .
\edm

\subsection{The interaction term}

We finally compute the interaction energy.
Using partial integration and the background equations
$\Diffbar^k(\alpha\bar{F}_{jk}) = 0$, it is convenient to
rewrite $E_{int}$ as
\bdm
E_{int} = -2G\int\left\{
    2L^{ij}\Trace\left( \bar{F}_i^{\; k}(\Diffbar{\cal E})_{kj} \right)
  + 2v^i l_i + 2\frac{\nabbar^k\alpha}{\alpha}\, v_k L
  + v^k \nabbar_k\tilde{A} \right\} \alpha\sqrt{\gbar}\,\diff\,x^3,
\edm
where $l_i = \alpha\nabbar^j( L_{ij}/\alpha )$
and $v^j = \Trace( \bar{F}^{kj} {\cal E}_k )$.

Inserting the above expansions for $L_{ij}$ and ${\cal E}_i$ and
integrating over the spherical variables, we obtain
\bea
E_{int} &=& \int
  \left( \sprod{V}{\bbA_{int}\p_\rho W} - \sprod{\bbA_{int}^T\p_\rho V}{W}\right. \nonumber\\
&& \left. \qquad\qquad +\, \sprod{V}{\bbS_{int} W} - \sprod{\bbb_{int}(\tilde{a})}{W} \right) \diff\rho,
\nonumber
\eea
where
\bdm
\bbA_{int} = \frac{u}{2} \left( \begin{array}{cccc}
  0 & 0 & -\sqrt{\frac{2}{3}} & 0 \\
  0 & 0 & \sqrt{\frac{2}{3}} & 0 \\
  1 & 0 & 0 & 0 \\
  0 & 0 & 0 & \sqrt{2}
  \end{array} \right),
\edm
\be
\bbS_{int} = \left( \begin{array}{cccc}
  \frac{\mu\gamma u}{\sqrt{3}} & -\frac{\mu\gamma^2 v}{\sqrt{3}}
  & -\frac{u_{,\rho}}{\sqrt{6}} - \sqrt{\frac{2}{3}}u \frac{(\gamma^2 r^3)_{,\rho}}{\gamma^2 r^3} & 0 \\
  \frac{2\mu\gamma u}{\sqrt{3}} & \frac{\mu\gamma^2 v}{\sqrt{3}}
  & \frac{7u_{,\rho}}{\sqrt{6}} + \sqrt{\frac{2}{3}}u \frac{(\gamma^2)_{,\rho}}{\gamma^2} & 0 \\
  \frac{3}{2}\,u_{,\rho }+ \frac{\gamma_{,\rho}}{\gamma}\, u
  & -\gamma wu - \frac{r}{\gamma}\!\left( \frac{\gamma^2}{r} v \right)_{\! ,\rho}
  & \frac{\mu\gamma u}{\sqrt{2}} & \sqrt{\frac{\lambda}{2}}\,\gamma u \\
  0 & -\sqrt{\lambda}\gamma^2 v & 0 & -\frac{u_{,\rho}}{\sqrt{2}}
  \end{array} \right),
\label{Eq-Sint}
\ee
and
\bdm
\bbb_{int}(\tilde{a}) = \frac{\alpha}{\sqrt{2}}\left( \mu u\, \tilde{a}, -\mu\gamma v\, \tilde{a}, \sqrt{2}r u\, \tilde{a}', 0 \right).
\edm

At this point, we remind the reader
that although the linearized electric YM field is
invariant under infinitesimal ${\mathfrak {su}}(2)$-{\it gauge\/}
transformations, it is not invariant under infinitesimal
{\it coordinate\/} transformations, see (\ref{Eq-YMGaugeFreedom}).
As a result, it is easy to show that the YM amplitudes $W$
transform according to
\bdm
W \mapsto W + \bbb_{int}(\xi),
\edm
where $f = \xi\, Y$.

\subsection{Summary}

Taking together the above results, the total energy functional becomes
\bea
E &=& \frac{1}{2} \int
  \Bigl( \sprod{\dot{U}}{\bbT\dot{U}} + \sprod{\p_\rho U}{\bbT\p_\rho U}
  + \sprod{U}{\bbA\p_\rho U} + \sprod{U}{\p_\rho\bbA U} \Bigr. \nonumber\\
&& \Bigl. \qquad + \,\sprod{U}{\bbS U} - 2\sprod{U}{\bbT\bbb(\tilde{a})_{grav}} \Bigr) \diff\rho,
\label{Eq-EnergyEYMLEven}
\eea
and the pulsation equations are
\be
\left( \bbT\p_t^{\, 2} - \bbT\p_\rho^{\, 2} + \bbA\p_\rho + \p_\rho\bbA
+ \bbS - \bbT\bbb(\tilde{a}) \right) U = 0,
\label{Eq-SymWaveEYMLEven}
\ee
where $U = (V,W)^T$, $\bbT = \diag(-1,1,1,1,1,1,1,1)$,
\bdm
\bbA = \left( \begin{array}{cc} 0 & \bbA_{int} \\
-\bbA^T_{int} & 0 \end{array} \right), \;\;\;
\bbS = \left( \begin{array}{cc} \bbS_{grav} & \bbS_{int} \\
\bbS^T_{int} & \bbS_{YM} \end{array} \right),
\edm
and
\bdm
\bbb(\tilde{a}) = \left( \bbb_{grav}(\tilde{a}), \bbb_{int}(\tilde{a}) \right).
\edm
Infinitesimal coordinate transformations are given by
\be
U \mapsto U + \bbb(\xi),
\label{Eq-EYMLGaugeTransf}
\ee
and the constraint equations are
\bea
0 &=& \frac{\alpha}{r^2}\p_{\rho}\left( \frac{r^2 p}{\alpha} \right)
   - \alpha r\p_{\rho} \left( \frac{t}{\alpha r} \right)
   - \frac{\sqrt{3}\mu}{2}\,\gamma\, q - \sqrt{\frac{3}{2}}\, u\, d,
\label{Eq-EYMLConsM1}\\
0 &=& \frac{\alpha}{r^2}\p_{\rho}\left( \frac{r^2 q}{\alpha} \right)
   - \frac{2\sqrt{3}\mu}{3}\,\gamma\, t - \frac{\sqrt{3}\mu}{3}\,\gamma\, p
   - \sqrt{\lambda}\gamma\, g - u\, b + \gamma v\, c,
\label{Eq-EYMLConsM2}\\
0 &=& \gamma\p_\rho\left( \frac{b}{\gamma} \right) + \gamma w c
   - \frac{\mu}{\sqrt{2}}\,\gamma\, d - \sqrt{\frac{\lambda}{2}}\,\gamma\, e.
\label{Eq-EYMLConsG}
\eea
Finally, as in our previous work, we assume that all perturbations
vanish at the boundary points, so that all boundary terms vanish.

\section{Special cases}
\label{Sect-5}

In this section, we discuss the perturbation equations for some
special cases. We first show that for $\ell = 0$, one recovers
the radial pulsation equation derived in \cite{W-Stable}.
Then, we introduce fully gauge-invariant amplitudes and discuss
the stability of the Schwarzschild-adS and the RN-adS solutions.
These examples, which are much simpler than the cases where the
gauge potential is effectively non-Abelian, indicate that
a topological stability analysis, as performed in the odd-parity
sector, is not possible in the even-parity sector.

\subsection{Radial perturbations}

For $\ell=0$, the only gravitational amplitudes are $t$ and $p$.
Since these are subject to the one-parameter family of
coordinate transformations (\ref{Eq-EYMLGaugeTransf}) and to the
constraint equation (\ref{Eq-EYMLConsM1}), one
expects that there are no dynamical gravitational modes. In fact,
one can show that for $\ell=0$, the only physical gravitational
perturbations correspond to the variation of the mass. This is
also clear in view of Birkhoff's theorem.
For the YM field, we have only the amplitude $d$ and the Gauss
constraint is void. Therefore, we expect to have an unconstrained
wave equation for the YM field when $\ell = 0$.
Looking at (\ref{Eq-EYMLGaugeTransf}) we see that we can construct
a gauge-invariant linear combination:
\bdm
\hat{d} = d + \frac{1}{\sqrt{6}} \frac{r u}{r_{,\rho}} (-t + p).
\edm
For computation, it is convenient to choose a gauge in which
$t=p$ since in this gauge, $\hat{d} = d$. Furthermore, in this gauge,
the only constraint equation yields the simple relation
\bdm
r_{,\rho }t = \frac{r u}{\sqrt{6}}\, d.
\edm
Using this, and taking the combination of the dynamical equations
which corresponds to the definition of $\hat{d}$, we obtain the
following wave equation for $d$:
\bea
&& \left( \p_t^{\, 2} - \p_\rho^{\, 2} \right)d
 + 2\frac{r u}{r_{,\rho}}\, u_{,\rho}\, d \nonumber\\
&& \qquad +\, \gamma^2\left[ 3w^2-1
 + \left( \frac{r u}{r_{,\rho}} \right)^2
   \left( -1 + 2N + \Lambda r^2 + G\frac{(w^2-1)^2}{r^2} \right)
   \right] d = 0. \nonumber
\eea
Finally, using $r u/r_{,\rho }= 2\sqrt{G} w_{,r}\,$,
$u^2 = 4G\gamma^2 N w_{,r}^2$
and the background YM equation
$r u_{,\rho }= 2\sqrt{G}\gamma^2 w(w^2-1) - u r_{,\rho }\,$, we have
\bea
&& \left( \p_t^{\, 2} - \p_\rho^{\, 2} + \gamma^2(3w^2-1) \right)d \nonumber\\
&& \qquad  +\, 4G\gamma^2\left[ 2ww_{,r}\frac{w^2-1}{r}
  - w_{,r}^2\left( 1 - \Lambda r^2 - G\frac{(w^2-1)^2}{r^2} \right) \right] d = 0,
\qquad\label{Eq-RadPertEq}
\eea
which agrees with the equation found in \cite{W-Stable}.
By replacing $d$ by $\hat{d}$, this equation acquires a
gauge-invariant meaning.
Using the estimate of section \ref{Sect-2} for $\Lambda
\rightarrow -\infty$, we see that the dominant terms in the
potential are $\gamma^2(3w^2-1)$ which is positive for solutions with
$w(r_h) > 1/\sqrt{3}$ and large $|\Lambda|$. This shows the
existence of solutions which are stable with respect to
even-parity radial perturbations \cite{W-Stable}.

\subsection{The gauge-invariant approach}

Motivated by the analysis above, one can try to
introduce gauge-invariant amplitudes for $\ell\geq 1$.
In order to do so, it turns out to be convenient to introduce the
amplitude
\bdm
\tau \equiv \frac{1}{\sqrt{6}} (-t+p) + \frac{r'}{\sqrt{2}\mu}\, q.
\edm
Using the transformation rules (\ref{Eq-VacGaugeFreedom}), we find
\bdm
\tau \mapsto \tau + \frac{\alpha^2}{r}\,f_{\tau }\,\xi,
\edm
where
\bdm
f_{\tau } \equiv \frac{\mu^2}{2} + r r' \frac{\alpha'}{\alpha} - N
 = \frac{\lambda}{2} + \frac{3m}{r} + Gw'^2 - \frac{G}{2}\frac{(w^2-1)^2}{r^2}\, .
\edm
The advantage of the new amplitude $\tau$ is that -- like the amplitude $g$ --
it transforms with an additive term which is algebraic in $\xi$. Provided
that $f_{\tau }$ has no zeros, one can use $\tau$ in order to construct
gauge-invariants. For example, we can construct
\be
\zeta = g - \sqrt{\frac{\mu^2\lambda}{2}}\frac{1}{f_{\tau }}\,\tau,
\label{Eq-zeta}
\ee
which is gauge-invariant for $\ell\geq 2$.
In fact, we will show below that for vacuum gravity, $\zeta$ satisfies
the Zerilli equation.

Using the momentum constraint equations (\ref{Eq-EYMLConsM1}) and
(\ref{Eq-EYMLConsM2}), we find the following constraint equation
\be
\gamma\p_\rho\left( \frac{\tau}{\gamma} \right) = \gamma(f_{\tau }
- 2Gw'^2)\frac{q}{\sqrt{2}\mu}
 + \sqrt{\frac{\lambda}{2\mu^2}}\frac{r_{,\rho}}{r}\, g + \frac{r'u}{\sqrt{2}\mu}\, b
 - \frac{1}{\sqrt{2}\mu}\frac{r_{,\rho}}{r}\, v\, c + \frac{u}{2}\, d.
\label{Eq-tauConstr}
\ee
In the gauge where $\tau = 0$, this yields an algebraic relation
between $q$, $g$, $b$, $c$ and $d$.

Next, we compute an evolution equation for $\tau$:
Taking the combination of the evolution equations which
corresponds to the definition of $\tau$, we arrive at the equation
\bea
0 & = &
 \boxdal\,\tau + \gamma^2\left(\mu^2 - \frac{6m}{r} + 5G\frac{(w^2-1)^2}{r^2}\right)\tau
   + \left(4\gamma_{,\rho }f_{\tau } - \frac{\alpha'}{\gamma} u^2 - \gamma u v w \right)\frac{q}{\sqrt{2}\mu}\nonumber\\
&& + \gamma^2 r'\left( -\frac{6m}{r} + 10Gw'^2 + 5G\frac{(w^2-1)^2}{r^2} \right)\frac{q}{\sqrt{2}\mu}
   \nonumber\\
&& + \frac{\sqrt{2\lambda}}{\mu}\gamma^2\left( 1-2N-r^2\Lambda - G\frac{(w^2-1)^2}{r^2}\right)g\nonumber\\
&& + \frac{1}{\sqrt{2}\mu}\left[ r'u\p_\rho b + 2r' u_{,\rho }b +
r'\frac{\gamma_{,\rho}}{\gamma}\, u b + \mu^2\gamma u\, b + 2\alpha r'' u b \right]\nonumber\\
&& - \frac{1}{\sqrt{2}\mu}\left[ \gamma w r' u -
\frac{rr'}{\gamma}\left( \frac{\gamma^2}{r}\, v \right)_{,\rho }- 2\alpha r'' \gamma v \right] c
 + \left( u_{,\rho }- \frac{1}{2}\frac{r_{,\rho}}{r}\, u \right)d\nonumber\\
&& + \frac{\sqrt{\lambda}}{2\mu}\frac{r_{,\rho}}{r}\, u\, e -
\frac{\alpha^2}{r}\, f_{\tau }\,\tilde{a} ,
\label{Eq-tau}
\eea
where we have defined
\bdm
\boxdal \equiv \left( \p_t^{\, 2} - \p_\rho^{\, 2} \right).
\edm
We have also used the momentum constraint equation (\ref{Eq-EYMLConsM2})
in order to eliminate first derivatives of $q$.
It seems clear from equation (\ref{Eq-tau}) that a formulation of the
evolution equations in terms of gauge-invariant quantities would
be quite messy. Worse than this, we have no reasons to expect
that the resulting equations have a symmetric potential.
Nevertheless, in some special cases where $w$ is constant,
the expressions simplify enough and one can find a fully gauge-invariant
description. This will be the subject of the next two subsections.

\subsection{Stability of the Schwarzschild-adS solution}

If $|w| = 1$ (i.e. the YM field is a pure gauge), we see that the
interaction terms $\bbA_{int}$ and $\bbS_{int}$ vanish, and the
gravitational and YM perturbations decouple from each other.
In order to investigate the gravitational sector, we adopt the
gauge-invariant formalism described in the previous subsection.
First, we note that for $\ell=1$, the constraint equation
(\ref{Eq-tauConstr}) yields, in the gauge $\tau=0$,
\bdm
q = 0.
\edm
Then, the evolution equation for $\tau$, (\ref{Eq-tau}), gives
$\tilde{a} = 0$. This shows that for $\ell=1$, the gravitational
sector is empty.

For $\ell\geq 2$, we introduce the gauge-invariant amplitude
\bdm
\zeta = g - \sqrt{\frac{\mu^2\lambda}{2}}\frac{1}{f_{\tau }}\,\tau,
\edm
where $f_{\tau } = \lambda/2 + 3m/r$.
We derive the equation for $\zeta$ in the gauge $\tau = 0$.
Taking the combination of the evolution equation for $\tau$, (\ref{Eq-tau}),
and the equation for $g$ corresponding to the definition of $\zeta$,
the terms involving $\tilde{a}$ cancel, and we obtain
\bdm
\boxdal\, g + \frac{\sqrt{\lambda}}{2}\,\gamma^2 r'\frac{6m}{r f}\, q + \left[ S_{33} -
 \gamma^2\frac{\lambda}{f_{\tau }}\left(1 - 2N - r^2\Lambda \right) \right]g = 0.
\edm
Next, the constraint equation (\ref{Eq-tauConstr}) yields
\bdm
q = -\sqrt{\lambda}\,\frac{r'}{f_{\tau }}\, g.
\edm
Using this, and $r_{,\rho }= N = 1 - 2m/r - \Lambda r^2/3$,
we eventually get the Zerilli equation \cite{Zerilli}
(with the presence of a cosmological constant)
\be
\left( \boxdal + V_Z \right)g = 0,
\label{Eq-ZerilliL}
\ee
where
\bdm
V_Z = N \frac{ \lambda^2 r^2 [ (\lambda+2)r + 6m ] + 12m^2 (3\lambda r + 6m - 2\Lambda r^3) }{(\lambda r + 6m)^2 r^3}\, .
\edm
If we replace $g$ by $\zeta$, this equation becomes manifestly
gauge-invariant. Since $V_Z$ is positive for all $r\in (r_h,\infty)$,
there are no gravitational instabilities.
At this point, let us also mention that equation
(\ref{Eq-ZerilliL}) and its odd-parity counterpart have already been
derived and used in order to compute the quasi-normal modes of
Schwarzschild-adS black holes (see Ref. \cite{CL1} and references
therein).

The YM perturbations describe a linearized YM field on a
Schwarzschild black hole. We will show later (see section \ref{Sect-6.2})
that those fields
cannot possess exponentially growing modes either.

\subsection{On the stability of the RN-adS solution}
\label{Sect-5.4}

When $w = 0$, we see that the equations decouple into two sets,
one set comprising the amplitudes $(t,p,q,g,c)$ (referred
to as Abelian
amplitudes in the following) and the other set comprising
$(b,d,e)$ (referred to as the non-Abelian amplitudes).

We start with the discussion of the Abelian amplitudes:
The relevant gauge-invariants are
\bea
\zeta &=& g - \sqrt{\frac{\mu^2\lambda}{2}}\frac{1}{f_{\tau }}\,
\tau, \nonumber\\
\psi &=& c + \frac{\mu}{\sqrt{2}}\,\frac{v}{f_{\tau }}\,\tau.\nonumber
\eea
(For $\ell=1$, $\zeta$ is void.)
Taking the combination of the evolution equations corresponding to
these amplitudes and eliminating $q$ using the constraint equation
(\ref{Eq-tauConstr}), i.e. using
\bdm
q = -\frac{r'}{f_{\tau }}\left[ \sqrt{\lambda} g - v c \right],
\edm
we get the following system of equations:
\be
\left( \boxdal + V_M \right)\left( \begin{array}{c} \zeta \\ \psi \end{array} \right) = 0,
\label{Eq-MoncriefL}
\ee
where the potential $V_M$ is symmetric and has the particular form
\bdm
V_M = N\left[ U + W\left( \begin{array}{cc} -3M & \sqrt{4G\lambda} \\
      \sqrt{4G\lambda} & 3M \end{array} \right) \right].
\edm
Using $N = 1 - 2M/r + G/r^2 - \Lambda r^2/3$ and
$2f_{\tau } = \lambda + 6M/r - 4G/r^2$, we can write the
functions $U$ and $W$ as (the fact that $f_{\tau} > 0$ for $r \geq r_h$ is
shown in Appendix~\ref{App-A})
\bea
U &=& \frac{1}{r^2(2f_{\tau })^2} \left[ \lambda^3
 + \left(2 + \frac{9M}{r} - \frac{4G}{r^2} \right)\lambda^2 \right. \nonumber\\
& &+ \left. 4\left( -\frac{2}{3}\,G\Lambda + \frac{3M}{r} + \frac{9M^2 + 2G}{r^2} - \frac{16GM}{r^3} + \frac{6G^2}{r^4} \right)\lambda \right. \nonumber\\
& &+ \left. 4\left( -3\Lambda M^2 + \frac{4G M\Lambda}{r} + \frac{9M^2 - \frac{8}{3}G^2\Lambda}{r^2} \right.\right.\nonumber\\
 && \left.\left.\quad\quad + \, \frac{9M^3}{r^3} - \frac{39G M^2}{r^4} + \frac{32M G^2}{r^5} - \frac{8G^3}{r^6} \right) \right], \nonumber\\
W &=& \frac{1}{r^3(2f_{\tau })^2}\left[ \lambda^2 + 4\lambda +
\frac{4M}{r}\left( 3 - \frac{3M}{r} + \frac{G}{r^2} \right) +
\frac{2\Lambda r^2}{3}\left( \frac{6M}{r} - \frac{8G}{r^2} \right) \right]. \nonumber
\eea
For $\Lambda = 0$, these equations reduce to the ones obtained by
Moncrief \cite{Moncrief} describing even-parity perturbations of a
RN black hole.
The fact that $V_M$ can be written as the sum of a function and a
function times a constant matrix allows the system to be
diagonalized. The absence of exponentially growing modes in this
system is discussed in Appendix \ref{App-A}.

For $\ell=0$, the non-Abelian modes are described by equation
(\ref{Eq-RadPertEq}) where one sets $w=0$:
\bdm
\left( \boxdal - \gamma^2 \right)d = 0.
\edm
In \cite{SW-Stable}, we have shown that this equation possesses
exponentially growing modes. Therefore, the RN-adS solution is
{\it unstable\/} with respect to non-Abelian odd- and even-parity
radial perturbations.

For $\ell\geq 1$, there are no instabilities. In order to see this,
we consider non-Abelian perturbations, which are described by
the system
\bea
0 & = &
 \boxdal\, b + \left[ \frac{\gamma_{,\rho\rho}}{\gamma} + \gamma^2(\lambda + 1)\right] b
     + \sqrt{2}\mu\gamma_{,\rho }\, d + \sqrt{2\lambda}\,\gamma_{,\rho}\, e ,
\nonumber\\
0 & = &
 \boxdal\, d + \sqrt{2}\mu\gamma_{,\rho}\, b + \gamma^2(\lambda + 1) d,
\nonumber\\
0 & = &
 \boxdal\, e + \sqrt{2\lambda}\,\gamma_{,\rho}\, b + \gamma^2(\lambda +
 1) e .
\nonumber
\eea
(For $\ell=1$, the equation for $e$ is not present.)
We also have the Gauss constraint,
\bdm
0 = \gamma\p_\rho\left(\frac{b}{\gamma}\right)
  - \frac{\mu}{\sqrt{2}}\,\gamma\, d - \sqrt{\frac{\lambda}{2}}\, \gamma\, e.
\edm
Using this constraint, we can eliminate both $d$ and $e$ in the evolution
equation for $b$. Defining $B = b/\gamma$, one obtains the
equation
\bdm
\boxdal\, B + \gamma^2(\lambda + 1) B = 0,
\edm
which has no unstable modes.
From the Gauss constraint, it follows that $\mu\, d + \sqrt{\lambda}\, e$
cannot grow exponentially in time either. Finally, we can find an
equation for $F \equiv d/\mu - e/\sqrt{\lambda}$:
\bdm
\boxdal\, F + \gamma^2(\lambda + 1) F = 0,
\edm
which is identical to the equation for $B$.
Stability for $\ell\geq 1$ will also follow from the
considerations made in section \ref{Sect-6.2}.

\section{The general case}
\label{Sect-6}

In this section we shall prove that both solitons and black holes
in which the gauge function $w$ has no zeros are stable
for $|\Lambda |$ sufficiently large.
Since the analysis required is rather involved, we shall
first outline the steps in the proof, before describing
the details. The sector with $\ell = 0$ has already been
discussed in \cite{W-Stable} and section \ref{Sect-5},
so we assume that $\ell \ge 1$ in the analysis below.

\subsection{Outline of stability proof}
\label{Sect-6.1}

In the last section, we have re-formulated the
perturbation equations on a Schwarzschild-adS and RN-adS
background in terms of fully gauge-invariant amplitudes.
Unfortunately, in the general case, this procedure is quite messy
and we have no guarantee that the resulting equations are going to be
suitable for analytic discussions.

In the following, we adopt a gauge in which $t=p$. Somehow, this
gauge has similar properties as $\tau=0$, but has the advantage of
being simpler computationally. In particular, one does not lose
the symmetry of the potential.
For $t=p$, the momentum constraint, Eq. (\ref{Eq-EYMLConsM1}),
reduces to an algebraic equation,
\be
\sqrt{6}\,\frac{r_{,\rho}}{r}\, t = \frac{\mu}{\sqrt{2}}\, \gamma q + u d,
\label{Eq-tConstr}
\ee
which permits one to re-express $t$ in terms of $q$ and $d$.
(Note that $r_{,\rho }= NS$ is positive everywhere outside the
horizon for black holes and positive everywhere for solitons.)
Introducing $t = p$ into the energy expression
(\ref{Eq-EnergyEYMLEven}), we see that there are no kinetic terms
involving $t$ (or $p$). In particular, the kinetic energy is
positive, and $t=p$ is non-dynamical.
The energy now has the form
\bea
E &=& \frac{1}{2} \int
  \Bigl( \sprod{\dot{\bar{U}}}{\dot{\bar{U}}} + \sprod{\p_\rho\bar{U}}{\p_\rho\bar{U}}
  + \sprod{\bar{U}}{\bar{\bbA}\p_\rho\bar{U}} + \sprod{\bar{U}}{\p_\rho\bar{\bbA}\bar{U}} \Bigr. \nonumber\\
&& \Bigl. \qquad + \,\sprod{\bar{U}}{\bar{\bbS}\bar{U}} - 2\sprod{\bar{U}}{\bar{\bbb}(\tilde{a})}
  + {\cal R} \Bigr) \diff\rho,
\label{Eq-gaugefixedE}
\eea
where $\bar{U} = (q,g,b,c,d,e)^T$ and where $\bar{\bbA}$,
$\bar{\bbS}$ and $\bar{\bbb}$ are obtained from $\bbA$, $\bbS$ and
$\bbb$ by removing rows and columns corresponding to $t$ and $p$.
The remainder terms ${\cal R}$ then are
\bea
{\cal R} &=& \left( S_{tt} + S_{11} + 2S_{t1} \right) t^2
+ 2\sqrt{12}\mu\gamma_{,\rho}\, t\, q\nonumber\\
        & &+ 2\sqrt{3}\mu\gamma u\, t\, b
+ 2\sqrt{6}\left( u_{,\rho }- u \frac{r_{,\rho}}{r} \right) t\, d
          + 2\left( b_t(\tilde{a}) - b_1(\tilde{a}) \right) t.
\label{Eq-remainderdef}
\eea
Variation of these equations with respect to $t$, $q$, $g$, $b$, $c$,
$d$, and $e$ yields the same equations as in (\ref{Eq-SymWaveEYMLEven})
but with $p=t$ and where the equations for $t$ and $p$ are summed,
resulting in an elliptic equation for $\tilde{a}$.

In order to discuss the stability of the system, we proceed as
follows:
We first give initial data for $t$, $p$, $q$, $g$, $b$, $c$, $d$, $e$
and their time derivatives such that the constraint equations
$t = p$, (\ref{Eq-EYMLConsM2}), (\ref{Eq-EYMLConsG}), (\ref{Eq-tConstr})
and their time derivatives are satisfied. Then, we evolve the data using the
hyperbolic evolution equations (\ref{Eq-SymWaveEYMLEven}).
In order to make sure that the gauge condition $t=p$ is preserved
during the evolution, we have to solve, at each time-step, the
elliptic equation for $\tilde{a}$ described above.
In \cite{SHB-PRD} we have shown that the constraint equations
propagate. In particular, this is the case for the constraint
(\ref{Eq-tConstr}).
Next, as we have shown in section \ref{Sect-3}, the terms which depend
on $\tilde{a}$ do not contribute to the energy $E$.
As a consequence, one can check that $E$ is conserved, i.e. $\p_t E = 0$,
provided that homogeneous Dirichlet boundary conditions are given
at the horizon and at infinity. Then, stability follows if we can show
that $E$ is positive. Indeed, if $E$ is positive, $L_{ij}$ and ${\cal E}_i$
cannot grow exponentially in time. In a gauge where the shift and the electric
potential are zero, the same follows for $\delta\gbar_{ij}$ and
$\delta\Abar_i$ since
\bea
&& \delta\gbar(t)_{ij} = \delta\gbar(0)_{ij} + \int\limits_0^t 2L(\tau)_{ij} \diff\tau, \nonumber\\
&& \delta\Abar(t)_i = \delta\Abar(0)_i - \int\limits_0^t {\cal E}_i(\tau) \diff\tau. \nonumber
\eea

Since the terms involving $\tilde{a}$ do not contribute, stability
will hold if the operator
\be
\bar{{\cal A}} \equiv -\p_\rho^{\, 2} + \bar{\bbA}\p_\rho
+ \p_\rho\bar{\bbA} + \bar{\bbS}
\label{Eq-stabop}
\ee
together with the remainder term ${\bar {\cal R}}$ is positive,
where
\bdm
{\bar {\cal {R}}}= {\cal {R}} -2\left( b_{t}({\tilde {a}})
-b_{1}({\tilde {a}}) \right) t.
\edm
From now on, we therefore ignore all terms dependent on
${\tilde  {a}}$, and shall consider only the energy functional
\be
{\tilde {E}}=E-\frac {1}{2} \int \left[
-2 \langle {\bar {U}},
{\bar {\bbb }}({\tilde {a}}) \rangle +2 \left(
b_{t}({\tilde {a}}) - b_{1}({\tilde {a}}) \right) t
\right] d\rho ,
\label{Eq-Etildedef}
\ee
since $E$ is positive (and the system stable) if ${\tilde {E}}$
is positive.

In the first stage of the proof (subsection \ref{Sect-6.2})
we show that the operator ${\hat {\cal {A}}}$
(which has the same form as ${\bar {\cal {A}}}$ (\ref{Eq-stabop})
but with slightly modified
matrices ${\bar {\bbA}}$ and ${\bar {\bbS}}$) can be written as
\bdm
{\hat {\cal {A}}} =
\bar{\bbB}^\dagger\bar{\bbB} + \bar{b}^T\bar{b},
\edm
where ${\bar {\bbB}}$ is a first order matrix differential operator and
${\bar {b}}$ is a vector.
This means that ${\hat {\cal {A}}}$ is a positive operator.

As a result of the factorization of ${\hat {\cal {A}}}$,
we now have a slightly modified remainder term ${\hat {\cal {R}}}$ given by
\bea
{\hat {\cal {R}}} &=&
\left( S_{tt} + S_{11} + 2S_{t1} \right) t^2
+ 2\sqrt{12}\mu\gamma_{,\rho}\, t\, q\nonumber\\
        & &   + 2\sqrt{3}\mu\gamma u\, t\, b + 2\sqrt{6} u_{,\rho }t\, d,
\label{Eq-remainder}
\eea
(ignoring the terms containing the lapse ${\tilde {a}}$).
Unfortunately, ${\hat {\cal {R}}}$ is not positive (even in the
Schwarzschild-adS case), so we do not immediately have that the energy
${\tilde {E}}$ is positive.
However,
in the next part of the stability proof (subsection \ref{Sect-6.3}),
we argue that this remainder term is of subleading order compared
to ${\hat {\cal {A}}}$ when $|\Lambda | \rightarrow \infty $.
This is not sufficient to give positivity of ${\tilde {E}}$ even for all
sufficiently large $|\Lambda |$, since the operator
${\hat {\cal {A}}}$ is
positive and not necessarily bounded away from zero (this means that
it may be that ${\hat {\cal {A}}}$ could be equal to zero for some
perturbations and then the subleading terms in ${\hat {\cal {R}}}$ would be
dominant, and may be negative).
From the fact that ${\hat {\cal {R}}}$ is subleading order as
$|\Lambda | \rightarrow \infty $,
we can however deduce that ${\tilde {E}}$ is positive for the precise value
$|\Lambda |=\infty $ (in a sense to be made exact in section \ref{Sect-6.3}).

It therefore remains to extend this stability for $|\Lambda |=\infty $
to sufficiently large values of $|\Lambda |$.
We will do this via an analyticity argument, using the
multi-dimensional equivalent of the nodal theorem \cite{AQ-Nodal}.
 The details of this are quite involved, and cover subsections
\ref{Sect-6.4}--\ref{Sect-6.6}.
First of all, as before we ignore the terms depending on
${\tilde  {a}}$ in the energy functional $E$ (\ref{Eq-gaugefixedE}),
and consider instead the modified energy functional
(\ref{Eq-Etildedef})
\bea
{\tilde {E}} &=& \frac{1}{2} \int
  \Bigl( \sprod{\dot{\bar{U}}}{\dot{\bar{U}}} + \sprod{\p_\rho\bar{U}}{\p_\rho\bar{U}}
  + \sprod{\bar{U}}{\bar{\bbA}\p_\rho\bar{U}} + \sprod{\bar{U}}{\p_\rho\bar{\bbA}\bar{U}} \Bigr. \nonumber\\
&& \Bigl. \qquad + \,\sprod{\bar{U}}{\bar{\bbS}\bar{U}}
  + {\bar {\cal {R}}} \Bigr) \diff\rho.
\nonumber
\eea
We also express $t$ (which is non-dynamical) in terms of $q$ and $d$
using the constraint equation (\ref{Eq-tConstr}).
The remainder term ${\bar {\cal {R}}}$ then takes the form:
\bdm
{\bar {\cal {R}}}= {\cal {S}}_{qq}q^{2}
+2{\cal {S}}_{qd}qd+{\cal {S}}_{dd}d^{2}
+2{\cal {S}}_{qb}qb+2{\cal {S}}_{bd}bd ,
\edm
where
\bea
{\cal {S}}_{qq} & = &
\frac {\mu ^{2}r^{2} \gamma ^{4}}{2r_{,\rho }^{2}}
\left[ -1+2N +\Lambda r^{2} +\frac {G(w^{2}-1)^{2}}{r^{2}} \right]
+\frac {2\mu ^{2}r\gamma _{,\rho }\gamma }{r_{,\rho }} ;
\nonumber \\
{\cal {S}}_{qd} & = &
\frac {\mu u r^{2} \gamma ^{3}}{{\sqrt {2}}r_{,\rho }^{2}}
\left[ -1+2N +\Lambda r^{2} +\frac {G(w^{2}-1)^{2}}{r^{2}} \right]
+\frac {{\sqrt {2}}\mu u r \gamma _{,\rho }}{r_{,\rho }}
+\frac {uu_{,\rho }r\gamma }{{\sqrt {2}}r_{,\rho }}
-\frac {u^{2}}{{\sqrt {2}}} ;
\nonumber \\
{\cal {S}}_{dd} & = &
\frac {u^{2}r^{2}\gamma ^{2}}{r_{,\rho }^{2}}
\left[ -1+2N +\Lambda r^{2} +\frac {G(w^{2}-1)^{2}}{r^{2}} \right]
+\frac {2uu_{,\rho }r}{r_{,\rho }} -2u^{2} ;
\nonumber \\
{\cal {S}}_{qb} & = &
\frac {\mu ^{2} u\alpha \gamma }{2r_{,\rho }} ;
\nonumber \\
{\cal {S}}_{bd} & = &
\frac {\mu u^{2}\alpha }{{\sqrt {2}}r_{,\rho }} .
\label{Eq-Sextrabits}
\eea
On varying ${\tilde {E}}$ with respect to the perturbations
$q$, $g$, $b$, $c$, $d$ and $e$, the remainder term
${\bar {\cal {R}}}$ will contribute additional terms to the
perturbation equations.
These additional terms are equivalent to adding the following
quantities to the pertinent parts of the potential ${\bar {S}}$:
add ${\cal {S}}_{qq}$ to $S_{22}$ in $\bbS _{grav}$ (\ref{Eq-Sgrav});
add ${\cal {S}}_{dd}$ to the $(dd)$-entry in
$\bbS _{YM}$ (\ref{Eq-S_YM}); finally, add
${\cal {S}}_{qd}$, ${\cal {S}}_{qb}$ and ${\cal {S}}_{bd}$
to the corresponding entries in $\bbS _{int}$ (\ref{Eq-Sint}).

There is one additional complication.
For both black hole and soliton solutions, it is found to be
necessary to study the perturbation equations arising not
from the variation of the whole of ${\tilde {E}}$, but, instead
we shall vary only
\bdm
E_{{\cal {P}}}={\tilde {E}}-{\cal {P}} ,
\edm
where ${\cal {P}}$ is manifestly positive.
Positivity of ${\tilde {E}}$ is automatic if $E_{{\cal {P}}}$
is positive since we shall always have ${\cal {P}}>0$.
Further, we can deduce that $E_{\cal {P}}\ge 0$ if the
perturbation equations derived from varying $E_{\cal {P}}$
have no unstable modes.
The splitting of ${\tilde {E}}$ in this form will
be seen to be necessary in section \ref{Sect-6.5} in
order to satisfy the technical conditions for the
application of the multi-dimensional nodal theorem.

In subsection \ref{Sect-6.4} we write the equations
derived from the variation of $E_{{\cal {P}}}$
in the form
\be
{\cal {O}} {\bar {U}} =
-\frac {d}{dR} \left( \bbK _{1}(R) \frac {d}{dR} {\bar {U}} \right) +
\left( \frac {\bbL (R)}{R^{2}} + \bbM (R) \right) {\bar {U}} =
- \bbK _{2}(R) \partial _{t}^{2} {\bar {U}} .
\label{Eq-nodalform}
\ee
In equation (\ref{Eq-nodalform}), the variable $R$ depends on
whether we are considering solitonic or black hole solutions and is
given in subsection \ref{Sect-6.5}, and
$\bbK _{1}$, $\bbK _{2}$, $\bbL$, and $\bbM$ are real
symmetric matrices depending on $R$.
We show, in subsection \ref{Sect-6.5}, that $\bbK _{1}$ is
uniformly positive definite, $\bbK _{2}$ is
positive definite,  $\bbL (0)$ is
non-negative and that all four matrices are smooth and uniformly
bounded for all $R$ in the interval $[0,\infty )$.
The crux of the stability proof is to show that the operator
${\cal {O}}$ has no negative eigenvalues for all $|\Lambda |$
sufficiently large, given that we have already proved that there are
no negative eigenvalues if $|\Lambda |=\infty $.
Since the matrix $\bbK _{2}$ is positive definite, stability follows
if the operator ${\cal {O}}$ has no negative eigenvalues.

Before applying the multi-dimensional nodal theorem \cite{AQ-Nodal},
we note that there are additional criteria required of the
matrices $\bbL $ and $\bbM $, which ensure that the essential
spectrum of the operator ${\cal {O}}$ equals $[0,\infty )$,
and that the number of bound states is finite.
Sufficient criteria are \cite{AQ-Nodal}:
$\bbM \rightarrow 0$ as $R\rightarrow \infty $; and
$R^{2} \bbM + \bbL \ge -\beta _{c}$ for some
$\beta _{c}<1/4$ and all sufficiently large $R$.
We will show, in section \ref{Sect-6.5}, that these additional
criteria are also satisfied by our matrices $\bbL $ and
$\bbM $.

The multi-dimensional nodal theorem \cite{AQ-Nodal} gives a
method for determining the number of negative eigenvalues of
the differential operator ${\cal {O}}$, as follows.
Fix $6$ (since we have a $6$-dimensional system)
linearly independent real $6$-vectors $e_{1},\ldots , e_{6}$,
and $a_{1}>0$.
Denote by
${\cal {U}}_{a_{1}}=[u_{1},\ldots , u_{6}]$ the $(6\times 6)$-matrix
whose columns are the solutions of the 6 initial value problems
\bdm
{\cal {O}} u_{j} =0, \qquad a_{1}<R<\infty , \qquad u_{j}(a_{1})=0,
\qquad \frac {du_{j}}{dR}(a_{1})=e_{j},
\qquad j=1,\ldots , 6.
\edm
Then the nodal theorem reads \cite{AQ-Nodal}:

\smallskip
{\bf{Theorem:}}
{\em {If $a_{1}>0$ is sufficiently small and
$a_{2}>a_{1}$ is sufficiently large, the
number of zeros (counted with multiplicities) in the interval
$(a_{1},a_{2})$
of the function
${\mathfrak {F}}:=R\mapsto \det {\cal {U}}_{a_{1}}(R)$ equals the number of
negative eigenvalues of ${\cal {O}}$ (counted with multiplicities).}}

\smallskip
At this point it should be stressed that the function
${\mathfrak {F}}$ must, by definition, vanish at $a_{1}$.
However, we already know that, when $|\Lambda |=\infty $,
the function ${\mathfrak {F}}$ has no zeros in the (open) interval
$(a_{1},a_{2})$, for sufficiently small $a_{1}$.
In subsection \ref{Sect-6.6} we show that each of the $u_{j}$ (and
hence ${\mathfrak {F}}$) are analytic in $\Lambda $ and $a_{1}$
in a neighbourhood of
$\Lambda =-\infty $, and for all $a_{1}>0$.
Therefore ${\mathfrak {F}}$
has no zeros in the interval $(a_{1},a_{2})$
for all $|\Lambda |$ sufficiently large (and all
sufficiently small $a_{1}$).
Therefore ${\cal {O}}$ has no negative eigenvalues for all
$|\Lambda |$ sufficiently large, and we have proved stability,
since the matrix $\bbK _{2}$ on the right-hand side of
(\ref{Eq-nodalform}) is positive definite.

\subsection{Factorization of a subsystem}
\label{Sect-6.2}

Having outlined the stability proof, in this section
we perform the first stage of this proof, by showing that
the operator ${\hat {\cal {A}}}$ is positive,
where ${\hat {\cal {A}}}$ has the same form
as the operator ${\bar {\cal {A}}}$ (\ref{Eq-stabop})
but with slightly modified matrices ${\bar {\bbA }}$ and
${\bar {\bbS }}$.

Firstly, we note that the operator
$\bar{{\cal A}}$ (\ref{Eq-stabop})
is very similar to the corresponding operator in the
odd-parity sector. This resemblance becomes even more striking if
we use the constraint equation (\ref{Eq-tConstr}) and replace the
term $-2\sqrt{6} u \frac{r_{,\rho}}{r}\, t\, d$ on the right-hand side
of equation (\ref{Eq-remainderdef}) by
\bdm
-\sqrt{2}\mu\gamma u\, q\, d - 2u^2 d^2.
\edm
Redefining $\bar{U} = (q,-b, g,-c,d,-e)^T$,
the energy functional ${\tilde {E}}$ (\ref{Eq-Etildedef})
then has the same form as before,
but where now
\bdm
\bar{\bbA} = \left( \begin{array}{cc} \bbA_1 & 0 \\ 0 & \bbA_2 \end{array} \right), \;\;\;
\bar{\bbS} = \left( \begin{array}{cc} \bbS_1 & \bbS_t^T \\ \bbS_t & \bbS_2 \end{array} \right),
\edm
with
\bea
&& \bbA_1 = -\frac{u}{2}\left( \begin{array}{cc} 0 & 1 \\ -1 & 0 \end{array} \right), \;\;\;
   \bbA_2 = -\frac{u}{\sqrt{2}}\left( \begin{array}{cccc} 0 & 0 & 0 & 1 \\ 0 & 0 & 0 & 0 \\
     0 & 0 & 0 & 0 \\ -1 & 0 & 0 & 0 \end{array} \right), \nonumber\\
&& \bbS_1 = \left( \begin{array}{cc}
   \frac{r}{\gamma}\left( \frac{\gamma}{r} \right)_{,\rho\rho} + \gamma^2(\lambda + v^2) & sym. \\
   -\frac{3}{2} u_{,\rho} - u \frac{\gamma_{,\rho}}{\gamma}
   & \frac{\gamma_{,\rho\rho}}{\gamma} + \gamma^2(\lambda + f) + u^2
   \end{array} \right),
\nonumber\\
&& \bbS_t = \left( \begin{array}{cc}
   2\sqrt{\lambda}\,\gamma_{,\rho }& 0 \\
   \frac{r}{\gamma}\left[ \frac{\gamma^2}{r} v \right]_{,\rho} + \gamma w u
   & -2(\gamma w)_{,\rho }- \gamma\, u v \\
   -\frac{\mu}{\sqrt{2}}\,\gamma\, u & -\sqrt{2}\mu \gamma_{,\rho }\\
   -\sqrt{\frac{\lambda}{2}}\,\gamma\, u & \sqrt{2\lambda}
   \gamma_{,\rho }
\end{array} \right),
\nonumber\\
&& \bbS_2 = \left( \begin{array}{cccc}
   \frac{r_{,\rho\rho}}{r} + \lambda\gamma^2 + u^2 & sym. & sym. & sym. \\
   \sqrt{\lambda}\gamma^2 v & \gamma^2(\lambda + 2f + v^2) & sym. & sym. \\
   0 & -\sqrt{2}\mu\gamma^2 w & \gamma^2(\lambda - 2 + 3f) & sym. \\
   \frac{u_{,\rho}}{\sqrt{2}} & -\sqrt{2\lambda}\gamma^2 w & 0 & \gamma^2(\mu^2 - f) \end{array} \right),
\nonumber
\eea
and
\bea
{\hat {\cal R}} &=&
\left( S_{tt} + S_{11} + 2S_{t1} \right) t^2
+ 2\sqrt{12}\mu\gamma_{,\rho}\, t\, q\nonumber\\
       & &  + 2\sqrt{3}\mu\gamma u\, t\, b + 2\sqrt{6} u_{,\rho }t\, d.
\nonumber
\label{Eq-remainderhat}
\eea
Comparing this with the corresponding expressions in the odd-parity sector
(see Eq. (8) of Ref. \cite{SW-Stable}) with $a=0$, we see that the only
difference lies in the sign of $(\bbS_2)_{23} = (\bbS_2)_{32}$.
It is also interesting to observe that in terms of the new variables,
the constraint equations (\ref{Eq-EYMLConsM2}) and (\ref{Eq-EYMLConsG})
are
\bea
0 &=& \frac{\gamma}{r}\p_{\rho}\left( \frac{r}{\gamma}\bar{q} \right)
   + u\,\bar{b} - \sqrt{\lambda}\gamma\, \bar{g}  - \gamma v\, \bar{c}
   - \sqrt{3}\,\mu\gamma\, t
\nonumber\\
0 &=& \gamma\p_\rho\left( \frac{1}{\gamma}\,\bar{b} \right) + \gamma w\bar{c}
   + \frac{\mu}{\sqrt{2}}\,\gamma\,\bar{d} - \sqrt{\frac{\lambda}{2}}\,\gamma\,\bar{e},
\nonumber
\eea
which, apart from the term involving $t$, are exactly the same as
the constraint equations for $h$ and $b$ in the odd-parity sector (see
Eq. (9) of Ref. \cite{SW-Stable}).

Therefore, if $\bbP$ denotes the projector from the even-parity
amplitudes $\bar{U} = (\bar{q},\bar{b},\bar{g},\bar{c},\bar{d},\bar{e})$
to the odd-parity amplitudes $U_{odd} = (h,a,b,k,c,d,e)$ defined in
Ref. \cite{SW-Stable},
\bdm
\bbP: \bar{U} \mapsto (\bar{q},0,\bar{b},\bar{g},\bar{c},\bar{d},\bar{e}),
\edm
and if $\bbE_{\bar{c}\bar{d}}$ is the matrix whose entry in the
column and row corresponding to $\bar{c}$ and $\bar{d}$,
respectively, is $1$ while all other entries are $0$, we have
\bdm
{\hat {\cal {A}}} = \bbP^T \bbB^\dagger\bbB\bbP
  - 2\sqrt{2}\mu\gamma^2 w\left(\bbE_{\bar{c}\bar{d}} + \bbE_{\bar{d}\bar{c}}\right),
\edm
where ${\hat {\cal {A}}}$ has the same form
as the operator ${\bar {\cal {A}}}$ (\ref{Eq-stabop}) but
with the matrices ${\bar {\bbA }}$ and ${\bar {\bbS }}$
replaced by their modified forms above.
The operator $\bbB$ and its adjoint $\bbB^\dagger$ were found
in \cite{SW-Stable}
($\bbB$ is also given in Appendix \ref{App-B} for completeness).
Now, introducing also
\bdm
\bbQ: U_{odd} = (h,a,b,k,c,d,e) \mapsto (0,a,0,0,0,0,0)
\edm
such that $\identy_{odd} = \bbP\bbP^T + \bbQ\bbQ^T$, and noticing
that
\bdm
\bbQ^T\bbB\bbP = \gamma(0,1,0,0,0,0,0)^T(0,0,0,-\mu,-\sqrt{2}w, 0),
\edm
we find
\bea
{\hat {\cal {A}}} &=&
(\bbP^T\bbB^\dagger\bbP)(\bbP^T\bbB\bbP)
 + \gamma^2\left( \mu^2\bbE_{\bar{c}\bar{c}}
- \sqrt{2}\mu w\bbE_{\bar{c}\bar{d}}
   -\sqrt{2}\mu w \bbE_{\bar{d}\bar{c}} + 2w^2\bbE_{\bar{d}\bar{d}}\right)
\nonumber\\
   &=& \bar{\bbB}^\dagger\bar{\bbB} + \bar{b}^T\bar{b},
\label{Eq-EvenFact}
\eea
where $\bar{\bbB} \equiv \bbP^T\bbB\bbP$ and
$\bar{b} = \gamma(0,0,0,-\mu,+\sqrt{2} w,0)$.
This shows that the operator ${\hat {\cal {A}}}$ is positive.
Note that this factorization also works when $\ell=1$, since
in that case the amplitudes $(\bar{q}, \bar{b}, \bar{c}, \bar{d})$
decouple from the amplitudes $(\bar{g},\bar{e})$ so that the above
result is still correct when $\bar{g}$ and $\bar{e}$ are removed.

As an application of (\ref{Eq-EvenFact}), consider the
Schwarzschild-adS case, $|w|=1$: It can be checked from the
explicit expressions in  Appendix \ref{App-B} that the YM
perturbations $(\bar{b},\bar{c},\bar{d},\bar{e})$ decouple from
the gravitational perturbations. Since the remainder term
$\hat{{\cal R}}$ does not depend on the YM perturbations in that
case, (\ref{Eq-EvenFact}) shows that the YM perturbations are
stable. Similarly, in the RN-adS case, the non-Abelian amplitudes
$(\bar{b},\bar{d},\bar{e})$ decouple, and (\ref{Eq-EvenFact})
shows that there are no instabilities. In contrast,  the
stability of the gravitational perturbations does not follow from
(\ref{Eq-EvenFact}) since in that case, the remainder term ${\hat {\cal {R}}}$
has to be taken into account. As a matter of fact, it turns out that
${\hat {\cal {R}}}$ is not positive even in the
Schwarzschild-adS case, and one needs to derive the master equations
(\ref{Eq-ZerilliL}) and (\ref{Eq-MoncriefL}) in order to prove
stability. In this sense, the gravitational perturbations are
responsible for the stability not being topological in the
even-parity sector!

The main idea in the next section will be to show that
${\hat {\cal {R}}}$ is negligible when $|\Lambda|$ is large enough and
to argue that in that case, the energy is still positive.

\subsection{Stability for infinite cosmological constant}
\label{Sect-6.3}

We will now argue that the remainder term
${\hat {\cal {R}}}$ (\ref{Eq-remainder}) given in the
previous subsection is sub-leading order as
$|\Lambda |\rightarrow \infty $.
In order to estimate carefully the magnitudes of all terms in our
system of equations, we firstly need to non-dimensionalize all
quantities.
All the perturbations, $q$, $g$, $b$, $c$, $d$ and $e$ have dimensions
of length. Bearing in mind that Newton's constant $G$ has
dimension $length^2$, we may write, for example,
\bdm
q = {\hat {q}}\, \ell_G
\edm
and similarly for the other quantities, where $\ell_G = \sqrt{G}\,$.
Since our equations are linear, the factors of $\ell_G$ all cancel and
the equations for the hatted variables are the same as those for the
unhatted variables.
Therefore we shall not distinguish in the following between the
dimensionless or dimensionful perturbations.

The time and radial variables are rescaled as:
\bdm
t = {\hat {t}}\,\ell_G\, , \qquad
\rho = {\hat {\rho }}\,\ell_G\, , \qquad
r = {\hat {r}}\,\ell_G\, .
\edm
The gauge function $w$ and all metric quantities are dimensionless.
The remaining dimensionful quantities are:
\bdm
r_{h} = \hat{r}_{h}\,\ell_G\, ,\qquad
m(r) = {\hat {m}}\, \ell_G\, , \qquad
u = {\hat {u}}\,\ell_G^{-1} = 2\frac {w_{,{\hat {\rho }}}}{\hat {r}\,\ell_G} , \qquad
\gamma = {\hat {\gamma }} \ell_G^{-1}.
\edm
We also define the positive dimensionless parameter
\bdm
L = -\Lambda\ell_G^{\, 2}\, .
\edm
All other quantities in the pulsation equations are dimensionless.

We are interested in the limit in which $L$ is very large (and
positive), and so we can introduce another positive parameter $\xi =
L^{-1}$, so that, equivalently, we may consider small $\xi $ in a
neighbourhood of $\xi =0$.
It is worth recalling at this stage that it has been proved that, in
this limit, ${\hat {u}}\sim o\left( \xi^{-1/2} \right)$
(see \cite{W-Stable} and section \ref{Sect-2}).
Furthermore, we know that in this limit the metric function ${\hat {m}}$
behaves like $L$, and, in turn, this means that $N \sim L$.
We therefore define
\bdm
{\tilde {N}}= \xi N =L^{-1} N,
\edm
and accordingly,
\bdm
{\tilde {\gamma }}^{2}=\xi {\hat {\gamma }}^{2},
\edm
where ${\tilde {N}}$ and ${\tilde {\gamma }}$ will be finite
as $\xi \rightarrow 0$.
With these scalings,
the potential ${\bar {\bbS }}$ is of order $L^{2}$ as
$L\rightarrow \infty $ (since each derivative
with respect to ${\hat {\rho }}$ introduces a factor of $N$),
so in order to obtain non-trivial
equations we must rescale the space-time variables again, namely,
\bdm
{\hat {t}}=L^{-1} {\tilde {t}}=\xi {\tilde {t}},
\qquad {\hat {r}}=\xi {\tilde {r}},
\qquad {\hat {\rho }}=\xi {\tilde {\rho }}.
\edm
The pulsation equations now have the form
\be
0  =
\left[ \partial _{\tilde {t}}^{2} - \partial _{\tilde {\rho }}^{2}
+ {\tilde {\bbA}} \partial _{\tilde {\rho }} +
\partial _{\tilde {\rho }} {\tilde {\bbA}} + {\tilde {\bbS}}
-{\tilde {\bbb}}({\tilde {a}}) \right] {\bar {U}} .
\label{Eq-pulsetilde}
\ee
where
\bdm
{\bar {\bbA}}={\tilde {\bbA}} L, \qquad
{\bar {\bbS}}={\tilde {\bbS}} L^{2}, \qquad
{\bar {\bbb}}={\tilde {\bbb}} L^{2}.
\edm
We also define
\bdm
{\bar {\cal {A}}}={\tilde {\cal {A}}} L^{2} \qquad
{\cal {R}}={\tilde {\cal {R}}} L^{2}.
\edm
The pulsation equations now have a well-defined limit as $\xi \rightarrow 0$
(or,  equivalently, $L\rightarrow \infty $).

The remainder term ${\tilde {\cal {R}}}$ (where ${\cal {R}}$ is given
  by (\ref{Eq-remainderdef})) contains $t$ terms.
The variable $t$ is non-dynamical and given in terms of the
  perturbations $q$ and $d$ by the constraint (\ref{Eq-tConstr}),
which reads, in dimensionless variables:
\be
t=\frac {\xi {\tilde {r}}}{{\sqrt {6}}{\tilde {r}}_{,{\tilde {\rho }}}}
\left( \frac {\mu }{{\sqrt {2\xi }}} {\tilde {\gamma }} q
+ {\hat {u}} d  \right) .
\label{Eq-tConstrxi}
\ee
We therefore substitute for $t$ to give an expression for
${\tilde  {\cal {R}}}$ in terms of $q$, $d$ and $b$.
From (\ref{Eq-tConstrxi}),
each term in ${\tilde {\cal {R}}}$ contains a factor of $\xi ^{1/2}$,
whereas the operator ${\tilde {\cal {A}}}$ is $O(1)$ as
$\xi \rightarrow 0$.
Therefore, as $\xi \rightarrow 0$, the remainder term
${\tilde {\cal {R}}}$ is subleading order compared to the operator
${\tilde {\cal {A}}}$.

So far in this section we have included the terms
in the perturbation equations dependent on the lapse
${\tilde {a}}$.
As discussed in section \ref{Sect-6.1}, these terms do
not contribute to the energy functional $E$, and so
may be ignored in our study of stability.
Ignoring the ${\tilde {\bbb }}$ terms therefore, our
conclusions apply equally well to the (unscaled) quantities
${\bar {\cal {A}}}$ and ${\bar {\cal {R}}}$
(see subsection \ref{Sect-6.1} for the definitions of these), namely
that, as $\xi \rightarrow 0$, the remainder term
${\bar {\cal {R}}}$ is subleading order compared
to the operator ${\bar {\cal {A}}}$.
This can also be extended to the slightly modified operator
${\hat {\cal {A}}}$ constructed in subsection \ref{Sect-6.2},
and the correspondingly modified remainder term
${\hat {\cal {R}}}$ given in (\ref{Eq-remainder}).
Therefore the (modified) remainder term ${\hat {\cal {R}}}$
is subleading order compared to the (modified) operator
${\hat {\cal {A}}}$, as $L\rightarrow \infty $.

By the factorization in subsection \ref{Sect-6.2}, we know that
the operator ${\hat {\cal {A}}}$ is positive, and therefore we
have proved that the total energy functional giving rise to the
pulsation equations (\ref{Eq-pulsetilde}) is positive when $\xi =0$
(or, equivalently, $L=\infty $), so that the system is stable
for this value of $\xi $. Since it is possible for ${\hat {\cal {A}}}$
to be zero for some perturbations, we are not able
immediately to say that the system is stable for sufficiently
small $\xi $ since it may be the case that for non-zero $\xi $ the
remainder term ${\hat {\cal {R}}}$ (which is not positive) is
dominant, so that the energy is no longer positive. The remainder
of this section will be devoted to extending this stability to
sufficiently small $\xi$, using an analyticity argument.

\subsection{Transformation of the equations}
\label{Sect-6.4}

We now address the problem of extending our proof that
the system of pulsation equations is stable when $\Lambda =-\infty $
to sufficiently large (and negative) $\Lambda $.
To do this, we shall use an analyticity argument based on the
multi-dimensional nodal theorem \cite{AQ-Nodal}.

In order to apply the multi-dimensional nodal theorem \cite{AQ-Nodal},
we first need to cast our pulsation equations into the appropriate
form.
We begin with the pulsation equations in the form
\bdm
0 =
\left( \partial _{t}^{2} - \partial _{\rho }^{2} +
{\bar {\bbA }}\partial _{\rho } + \partial _{\rho }
{\bar {\bbA }}+ {\bar {\bbS }} \right)
{\bar {U}},
\edm
where we remind the reader that ${\bar {\bbA }}$ and
${\bar {\bbS }}$ are anti-symmetric
and symmetric matrices, respectively,
and ${\bar {U}}=(q,g,b,c,d,e)$.
Note that we are ignoring the terms in the perturbation equations
which contain the lapse, since these terms do not contribute to the
energy $E$.
Furthermore, we derive these equations by firstly substituting
for the variable $t=p$ using the momentum constraint
(\ref{Eq-tConstr}),
which gives the remainder term ${\bar {\cal {R}}}$
whose terms are given by (\ref{Eq-Sextrabits}) .
Finally, we vary not the whole modified energy ${\tilde {E}}$
(\ref{Eq-Etildedef}),
but rather $E_{{\cal {P}}}$, which is the modified energy
${\tilde {E}}$ minus a term ${\cal {P}}$ which is manifestly
positive.
This procedure is outlined in section \ref{Sect-6.1},
and the details for the particular cases of black hole
and soliton equilibrium solutions are given in section
\ref{Sect-6.5} below.

Define a new perturbation vector ${\bar {V}}$ by
\bdm
{\bar {U}}=\bbR {\bar {V}}
\edm
where $\bbR $ is a transformation matrix, which may depend on $\rho $
(but not $t$) and
contains no derivative operators.
Then the equation for ${\bar {V}}$ is
\be
0 =
\left[
\bbR ^{T} \bbR \partial _{t}^{2} - \partial _{\rho }
\bbR ^{T}  \bbR \partial _{\rho } + {}^{1}{\bar {\bbA }} \partial _{\rho }
+ \partial _{\rho }{}^{1} {\bar {\bbA}} + {}^{1}{\bar {\bbS }}
\right] {\bar {V}} ,
\label{Eq-V}
\ee
where the transformed matrices ${}^{1}{\bar {\bbA }}$
and ${}^{1}{\bar {\bbS }}$ are
\bea
{}^{1}{\bar {\bbA }} & = & \bbR ^{T} \bbA \bbR + \frac {1}{2} \left(
\bbR ^{T}_{,\rho }\bbR - \bbR ^{T} \bbR_{,\rho } \right) ,\nonumber \\
{}^{1}{\bar {\bbS}} & = & \bbR ^{T} \bbS \bbR + \bbR ^{T} \bbA \bbR_{,\rho }
-\bbR ^{T}_{,\rho } \bbA \bbR -\frac {1}{2}\left( \bbR ^{T} \bbR _{,\rho \rho }
+ \bbR ^{T}_{,\rho \rho } \bbR \right).
\nonumber
\eea
Since
\bdm
{\bar {\bbA }}= \left(
\begin{array}{cc}
0 & {\bar {\bbA }}_{int} \\
-{\bar {\bbA }}_{int}^{T} & 0
\end{array}
\right) ,
\edm
we set
\bdm
\bbR = \left(
\begin{array}{cc}
\bbI & 0 \\
\bbF ^{T} & \bbI
\end{array}
\right) ,
\edm
where $\bbF $ is a $2\times 4$ matrix.
If we choose $\bbF$ such that
\bdm
{\bar {\bbA }}_{int} = -\frac {1}{2} \bbF _{,\rho },
\edm
then ${}^{1}{\bar {\bbA}}=0$, and the pulsation equations simplify.
With this choice, the matrix $\bbF $ takes the form
\bdm
\bbF = {\cal {F}} \left(
\begin{array}{cccc}
1 & 0 & 0 & 0 \\
0 & 0 & 0 & {\sqrt {2}}
\end{array}
\right)
\edm
where the scalar function ${\cal {F}}(\rho )$ is given by
\be
{\cal {F}}(\rho) =
-2{\sqrt {G}} \int^\rho \frac {w_{,\rho }}{r} \, d\rho = -2 {\sqrt {G}}
\int^r \frac {w_{,r}}{r} \, dr .
\label{Eq-Fdef}
\ee
We write the transformed potential as:
\be
{}^{1}{\bar {\bbS }} = \bbR ^{T} \left(
\begin{array}{cc}
{}^{1}{\bar {\bbS }}_{grav} & {}^{1}{\bar {\bbS}} _{int} \\
{}^{1}{\bar {\bbS }}_{int} & {}^{1}{\bar {\bbS }}_{YM}
\end{array}
\right) \bbR ,
\label{Eq-Sbar}
\ee
where
\bea
{}^{1}{\bar {\bbS }}_{grav} & = & {\bar {\bbS }}_{grav}
- 4 {\bar {\bbA }}_{int} {\bar {\bbA }}_{int}^{T} ,
\nonumber \\
{}^{1}{\bar {\bbS }}_{int} & = & {\bar {\bbS }}_{int}
+ {\bar {\bbA }}_{int  \, ,\rho } ,
\nonumber \\
{}^{1}{\bar {\bbS }}_{YM} & = & {\bar {\bbS }}_{YM} .
\label{Eq-Sbarbits}
\eea

The transformation matrix $\bbR $ has all its eigenvalues
equal to unity, but it is not necessarily positive definite
because it may contain a Jordan block.
However, if we introduce the vector
$\bbX = (x_{1}, x_{2}, x_{3}, x_{4}, x_{5}, x_{6} )^{T}$,
then
\bea
\bbX ^{T} \bbR \bbX & = &
\left( x_{1}+ {\cal {F}}(\rho ) x_{3} \right) ^{2}
+ \left( x_{6}+ {\sqrt {2}} {\cal {F}}(\rho )
x_{2} \right) ^{2}
\nonumber
\\
& &
+ x_{4}^{2}+x_{5}^{2}
+\left( 1 -2{\cal {F}}^{2} (\rho ) \right) x_{2}^{2}
+\left( 1- {\cal {F}}^{2} (\rho ) \right) x_{3}^{2}.
\nonumber
\eea
Therefore $\bbR $ is positive definite if
${\cal {F}}^{2}(\rho )< 1/2$.
Since (see section 2 and \cite{W-Stable}),
$w_{,r}(r) \sim o(|\Lambda |^{-\frac {1}{2}})$ as
$|\Lambda |\rightarrow \infty $, for sufficiently large
$|\Lambda |$ (and an appropriate choice of integration
constant) we can make ${\cal {F}}$ as small as we like.
Therefore $\bbR $ is positive definite for sufficiently large
$|\Lambda |$.

We now make a further transformation by introducing a
diagonal matrix $\bbY $ (depending only on $\rho $ and not on $t$),
which commutes with $\bbR $.
In order to commute with $\bbR $, the elements of $\bbY $ must be such
that
\be
\bbY = \diag(y_{1}, y_{2}, y_{1}, y_{3}, y_{4}, y_{2}) .
\label{Eq-Ydef}
\ee
Defining ${\bar {W}}=\bbY ^{-1}{\bar {V}}$,
the pulsation equations (\ref{Eq-V}) now
take the form
\be
0 =
\bbY \bbR ^{T} \bbR \bbY \partial _{t}^{2} {\bar {W}}
-\partial _{\rho } \left( \bbY \bbR ^{T} \bbR \bbY \partial _{\rho }
{\bar {W}} \right)
+ {}^{2}{\bar {\bbS }} {\bar {W}},
\label{Eq-W}
\ee
where the new potential is
\bea
{}^{2}{\bar {\bbS }} & = & \bbY \left[ {}^{1}{\bar {\bbS }} - \partial _{\rho } \left(
\bbR ^{T} \bbR \right) \bbY ^{-1} \bbY _{,\rho } -
\bbR ^{T} \bbR \bbY ^{-1} \bbY _{,\rho \rho } \right] \bbY
\nonumber \\
 & = &
\bbY\bbR ^{T} \left( \begin{array}{cc}
{}^{2}{\bar {\bbS }}_{grav} & {}^{2}{\bar {\bbS }}_{int} \\
{}^{2}{\bar {\bbS }}_{int}^{T} & {}^{2}{\bar {\bbS }}_{YM}
\end{array}
\right) \bbR \bbY,
\label{Eq-Stilde}
\eea
and
\bea
{}^{2}{\bar {\bbS }}_{grav} & = &
{}^{1}{\bar {\bbS }}_{grav} - \bbY _{1}^{-1} \bbY _{1 \, ,\rho \rho } ,
\nonumber \\
{}^{2}{\bar {\bbS }}_{int} & = & {}^{1}{\bar {\bbS }}_{int} - \bbF _{,\rho }
\bbY _{2}^{-1} \bbY _{2 \, ,\rho } ,
\nonumber \\
{}^{2}{\bar {\bbS }}_{YM} & = & {}^{1}{\bar {\bbS }}_{YM} - \bbY _{2}^{-1}
\bbY _{2 \, ,\rho \rho } ,
\label{Eq-Stildebits}
\eea
and we have defined diagonal matrices $\bbY_{1}$ and $\bbY_{2}$,
which form the blocks of the matrix $\bbY $, as follows:
\bea
\bbY _{1} & = & \diag(y_{1},y_{2}),
\nonumber \\
\bbY _{2} & = & \diag(y_{1}, y_{3}, y_{4}, y_{2}) .
\nonumber
\eea
At the moment we leave the functions in the matrix $\bbY $ to be quite
general.
They will be fixed in the next subsection \ref{Sect-6.5}, when we
discuss the form of the equations in detail for black holes and
solitons in turn.

For the moment, we write the pulsation equations (\ref{Eq-W}) as
\be
-\bbK (\rho )\partial _{t}^{2} {\bar {W}} = -\partial _{\rho } \left(
\bbK (\rho ) \partial _{\rho } {\bar {W}} \right)
+ {}^{2}{\bar {\bbS }} {\bar {W}},
\label{Eq-transformed}
\ee
where
\bdm
\bbK = \bbY \bbR ^{T} \bbR \bbY .
\edm
Equation (\ref{Eq-transformed})
should be compared with the form (\ref{Eq-nodalform})
required for the application of the nodal theorem \cite{AQ-Nodal}.
The first issue is whether our coordinate $\rho $ has the
required domain $[0,\infty )$.
This will be addressed in the following subsection for
the black hole and soliton solutions.
For the time being, we assume that we can, if necessary,
change to a coordinate $R$ (which will be a function of $\rho $)
which does have the required range of values.
If
\be
\frac {dR}{d\rho } =Q,
\label{Eq-Qdef}
\ee
then the equations (\ref{Eq-transformed}) become
\bdm
-\bbK Q^{-1} \partial ^{2}_{t} {\bar {W}}
= -\frac {d}{dR} \left(
 \bbK Q \frac {d{\bar {W}}}{dR} \right)
+ {}^{2}{\bar {\bbS }} Q^{-1} {\bar {W}} ,
\edm
which is of the form (\ref{Eq-nodalform}), with
$\bbK _{1}=\bbK Q$ and $\bbK _{2}=\bbK Q^{-1}$.
In our work in the following section, the function $Q$ will
always be of one sign, and so we consider instead
(by multiplying throughout by -1 if necessary)
\be
- \bbK |Q|^{-1} \partial ^{2}_{t} {\bar {W}}
=-\frac {d}{dR} \left(
\bbK |Q| \frac {d{\bar {W}}}{dR} \right)
+{}^{2}{\bar {\bbS }}|Q|^{-1} {\bar {W}},
\label{Eq-transformed1}
\ee
so that the matrices $\bbK _{1}=\bbK |Q|$ and
$\bbK _{2}=\bbK |Q|^{-1}$ are positive definite.
As discussed in section \ref{Sect-6.1}, in order to prove stability
we require that $\bbK _{1}$ is
uniformly bounded and uniformly positive
definite, a condition which will be considered in the next subsection.
In order to apply the generalized multi-dimensional nodal theorem,
it remains to show that the matrix ${}^{2}{\bar {\bbS }}|Q|^{-1}$
can be written in the form
\be
{}^{2}{\bar {\bbS }} |Q|^{-1}
=\frac {1}{R^{2}} \bbL (R) + \bbM (R) ,
\label{Eq-nodalpotential}
\ee
where $\bbL $ and $\bbM $ are uniformly bounded matrices on
$R \in [0,\infty )$, and $\bbL (0)$ is non-negative.
Furthermore, we also require $\bbM \rightarrow 0$ as
$R \rightarrow \infty $, and $\bbM R^{2} +\bbL \ge -\beta _{c}$
for some $\beta _{c}< 1/4$ and all sufficiently large $R$.
This is the subject of the next subsection.

\subsection{The nodal theorem matrices}
\label{Sect-6.5}

Having cast our pulsation equations into a form
(\ref{Eq-transformed1}) which will allow us to apply the
(extended) multi-dimensional nodal theorem, we now
have to verify that the matrices in (\ref{Eq-transformed1}) do
in fact satisfy the conditions of the nodal theorem.
Namely, $\bbK |Q|$ and
$\bbK |Q|^{-1}$ must be uniformly bounded and
positive definite, and the potential ${}^{2}{\bar {\bbS }}|Q|^{-1}$
must have the form (\ref{Eq-nodalpotential}).
In addition, $\bbM \rightarrow 0$ as
$R \rightarrow \infty $, and $\bbM R^{2} +\bbL \ge -\beta _{c}$
for some $\beta _{c}< 1/4$ and all sufficiently large $R$,
which will be the case if
${}^{2}{\bar {\bbS }}|Q|^{-1}\rightarrow 0$ as
$R\rightarrow \infty $, and
${}^{2}{\bar {\bbS }}|Q|^{-1}R^{2}\ge -\beta _{c}$
for all sufficiently large $R$.
So far we have also left open the precise identity of
the coordinate $R$ for use in the application of the
nodal theorem.

In this subsection we shall address these issues.
The two cases, black hole  and soliton solutions, need to be
addressed separately because of the different boundary conditions
on the equilibrium solutions.
The black hole perturbation equations are more readily
cast into an appropriate form, so we deal with them first.

\subsubsection{Black holes}
\label{Sect-6.5.1}

For black hole equilibrium solutions, the usual ``tortoise''
coordinate $\rho $ has the domain $(-\infty , \rho _{max}]$,
tending to $-\infty $ close to the event horizon, and
approaching the finite value $\rho _{max }$ at infinity.
By a suitable choice of integration constant in the definition of
$\rho $, we may set $\rho _{max}$ to be equal to zero.
Therefore $\rho \in (-\infty , 0] $.
Our application of the multi-dimensional nodal theorem
requires a variable $R$ having domain $[0,\infty )$,
so we define
\bdm
R=-\rho .
\edm
It should be noted that $R=0$ corresponds now to infinity,
and $R\rightarrow \infty $ corresponds to the event horizon.

In this situation, the quantity $Q$ (see (\ref{Eq-Qdef})) is
equal to $-1$, and so $|Q|=1$.
We set the transformation matrix $\bbY $ (\ref{Eq-Ydef}) to be
equal to the identity,
so that the perturbation  equations are now of the form
(\ref{Eq-transformed1})
\bdm
-\bbK  \partial ^{2}_{t} {\bar {V}} = -\frac {d}{dR} \left(
 \bbK  \frac {dV}{dR} \right) + {}^{1}{\bar {\bbS }}  {\bar {V}} ,
\edm
where
\bdm
\bbK = \bbR ^{T} \bbR ,
\edm
and ${}^{1}{\bar {\bbS }}$ is given by (\ref{Eq-Sbar}).

The first necessary condition is that the matrix $\bbK$
is uniformly bounded and
uniformly positive definite on $R\in [0,\infty )$.
Recall that $\bbR $ depends on the function ${\cal {F}}(R)$,
defined by (\ref{Eq-Fdef}).
We are interested in those black holes for which the gauge function
$w$ has no zeros, in which case,
from the field equations, the function $w$ is a monotonic function
and so $w_{,r}$ is of one sign.
To see this, suppose that $w>0$.
First note that the field equations (\ref{eq-equil})
imply that $w$ cannot have a minimum in the interval $0<w<1$
nor a maximum if $w>1$ \cite{breit}.
Furthermore, a point of inflection is possible only if $w=1$,
and it was proven in \cite{SW-Stable} that for solutions in which
$w$ has no nodes, the function $w$ cannot cross $\pm 1$.
Taken together, these statements mean that $w$ can have no stationary
points, and this argument applies equally well to soliton
solutions or if $w<0$ everywhere.

Given the asymptotic behaviour of $w$ at infinity (see section
\ref{Sect-2}), in this case the integral defining ${\cal {F}}$
(\ref{Eq-Fdef}) is convergent
at infinity (which corresponds to $R=0$).
This means that the matrix $\bbR $ is uniformly bounded on
$R\in [0,\infty )$, so that $\bbK $ is also uniformly bounded.
It is also straightforward to show that $\bbK $ is uniformly positive
definite on $R\in [0,\infty )$ in this case.

Since the matrix $\bbR $ is uniformly bounded
and positive definite,
to show that the potential
${}^{1}{\bar {\bbS }}$ can be written in the form
(\ref{Eq-nodalpotential}),
it is sufficient to show that this is the case for the potential
matrix
\bdm
\left( \begin{array}{cc}
{}^{1}{\bar {\bbS }}_{grav} & {}^{1}{\bar {\bbS }}_{int} \\
{}^{1}{\bar {\bbS }}_{int}^{T} & {}^{1}{\bar {\bbS }}_{YM}
\end{array}
\right) ,
\edm
whose components are given in (\ref{Eq-Sbarbits}).
Remember that $R=0$ (which is where the potential can be singular)
corresponds to infinity.
It is simplest to consider each part of the potential
separately.

Firstly, from (\ref{Eq-Sbarbits}), there are terms
in the potential
proportional to ${\bar {\bbA }}_{int}$ and/or
its derivative with respect to $\rho $.
These depend only on the gauge function $u$ (\ref{Eq-udef})
and its derivative, respectively.
Both these functions are regular everywhere outside the
event horizon and uniformly bounded for $R\in [0,\infty )$.
Therefore those parts of the potential containing
${\bar {\bbA }}_{int}$
or its derivative may be included in the matrix $\bbM $
(\ref{Eq-nodalpotential}).

It now remains to consider only the terms
${\bar {\bbS }}_{grav}$,
${\bar {\bbS }}_{int}$ and ${\bar {\bbS }}_{YM}$,
remembering to include the additional contributions
from the remainder term (\ref{Eq-Sextrabits}).
Careful analysis of each component in these matrices reveals
that all components are regular and uniformly bounded everywhere
outside the event horizon
(so that they may be included in the matrix $\bbM $)
 with the sole exception
of the quantity
\bdm
\frac {r_{,\rho \rho }}{r}
\edm
which arises in the $S_{33}$ component of the ${\bar {\bbS }}_{grav}$
matrix (see (\ref{Eq-Sgrav})).
At infinity ($R\rightarrow 0$), we have
\bdm
\frac {r_{,\rho \rho }}{r} = \frac {2}{R^{2}} + O(1).
\edm
We therefore define a matrix $\bbL $ which has
as its only non-zero element
\bdm
\bbL _{33} = \frac {R^{2}r_{,\rho \rho }}{r} .
\edm
Near the event horizon $R\sim -\log (r-r_{h})\rightarrow \infty$, and
so $\bbL _{33} \rightarrow 0$.
Thus the matrix $\bbL $ is symmetric, regular and
uniformly bounded on $R\in [0,\infty )$.
In addition, $\bbL _{33}(0)=2$, so that the
matrix $\bbL (0)$ is non-negative, as required.
We comment here that a crude estimate of the behaviour
of
\bdm
\frac {r^{2}}{\alpha } \left( \frac {\alpha }{r^{2}} \right)
_{,\rho \rho }
\edm
which arises in the $S_{22}$ component of ${\bar {\bbS }}_{grav}$
suggests that this is divergent at infinity.
However, a precise calculation, using the asymptotic
forms of the metric functions (section \ref{Sect-2})
shows that in fact this function is regular everywhere outside the
event horizon.

Having written the potential ${}^{1}{\bar {\bbS }}$
in the required form (\ref{Eq-nodalpotential}), it remains
to show that this potential vanishes sufficiently quickly
as $R\rightarrow \infty $, that is,
\bdm
{}^{1}{\bar {\bbS }}R^{2} \ge -\beta _{c}
\edm
for all sufficiently large $R$, where $\beta _{c}$ is a constant
which is less than $1/4$.
Recall that $R\rightarrow \infty $ corresponds to the event horizon
of the black hole.
As the event horizon is approached, all terms in the potential
${}^{1}{\bar {\bbS }}$ tend to zero at least as rapidly
as $(r-r_{h})^{1/2}$, with the exception of the following terms:
\bea
{\cal {S}}_{qq}& = & \frac {\mu ^{2}S^{2}(r_{h})N_{,r}(r_{h})}{2r_{h}}
+O(r-r_{h});
\nonumber \\
S_{22} & = & \frac {1}{4} S^{2}(r_{h}) N_{,r}^{2}(r_{h})
+O(r-r_{h});
\nonumber
\eea
where ${\cal {S}}_{qq}$ is the extra contribution to $S_{22}$
in $\bbS _{grav}$ (\ref{Eq-Sgrav}) given in (\ref{Eq-Sextrabits}).
In addition, the $(bb)$-component of $\bbS _{YM}$ (\ref{Eq-S_YM})
has the same behaviour as $S_{22}$.
Therefore there are two entries in the potential matrix which
approach a positive constant as $R\rightarrow \infty $ (since
$N_{,r}(r_{h})>0$).
We deal with these by defining a manifestly positive quantity
${\cal {P}}$:
\bdm
{\cal {P}}=
\left[  \frac {\mu ^{2}S^{2}(r_{h})N_{,r}(r_{h})}{2r_{h}} +
\frac {1}{4} S^{2}(r_{h}) N_{,r}^{2}(r_{h}) \right] q^{2}
+ \frac {1}{4} S^{2}(r_{h}) N_{,r}^{2}(r_{h}) b^{2}
>0
\edm
which we subtract from the modified energy functional
${\tilde {E}}$ before varying
$E_{{\cal {P}}}= {\tilde {E}}-{\cal {P}}$
to yield the perturbation equations.
These equations are identical to the previous ones
apart from two terms in the untransformed potential ${\bar {\bbS }}$:
the $(qq)$-term in ${\bar {\bbS }}_{grav}$, which has the
positive constant
\bdm
\left[  \frac {\mu ^{2}S^{2}(r_{h})N_{,r}(r_{h})}{2r_{h}} +
\frac {1}{4} S^{2}(r_{h}) N_{,r}^{2}(r_{h}) \right]
\edm
subtracted, and the $(bb)$-term in ${\bar {\bbS }}_{YM}$,
which has the positive constant
\bdm
\frac {1}{4} S^{2}(r_{h}) N_{,r}^{2}(r_{h})
\edm
subtracted.
This does not affect
any of the analysis so far in this subsection, in particular
the definition of the matrix $\bbL $ is unchanged, and the
matrix $\bbM $ simply has appropriate positive constants
subtracted from its $(qq)$ and $(bb)$ entries.
Therefore both $\bbL$ and $\bbM$ remain uniformly bounded
for all $R\in [0,\infty )$.

However, now all entries in the potential ${}^{1}{\bar {\bbS }}$
vanish at least as quickly as $(r-r_{h})^{1/2}$ as
$r\rightarrow r_{h}$ (or, equivalently, $R\rightarrow \infty $).
Since, near the event horizon, $R\sim -\log (r-r_{h})$,
this means that the potential is vanishing like $e^{-R/2}$ as
$R\rightarrow \infty $.
Therefore ${}^{1}{\bar {\bbS }} R^{2}\rightarrow 0$ as
$R\rightarrow \infty $, so it is certainly the case that
\bdm
{}^{1}{\bar {\bbS }} R^{2} \ge -\beta _{c}
\edm
where $\beta _{c}<1/4$ for all sufficiently large $R$.

Therefore, for black hole equilibrium solutions, we have constructed
matrices satisfying the conditions necessary for the application
of the multi-dimensional nodal theorem.

\subsubsection{Solitons}
\label{Sect-6.5.2}

The corresponding analysis for soliton solutions is
rather more complex.
We shall begin by defining
${\cal {P}}=(\gamma ^{4}\mu ^{2}r^{2}/2r_{,\rho }^{2})
q^{2}$ as it is manifestly positive,
and consider the perturbation equations derived from the
variation of $E_{{\cal {P}}}={\tilde {E}}-{\cal {P}}$
(as outlined in section \ref{Sect-6.1}).
This makes some of the algebra later in this subsection
more tractable.
We can prove stability for soliton solutions
 if these pulsation equations have no unstable modes.
Therefore, for the pulsation equations for solitons,
we remove a term
\bdm
\frac {\mu ^{2} \gamma ^{4} r^{2}}{2r_{,\rho }^{2}}
\edm
from $S_{22}$ in ${\bar {\bbS }}_{grav}$.
All other terms in the potential are unchanged.

For soliton solutions, the ``tortoise'' co-ordinate $\rho $
has a finite range of values, so we shall instead begin with the
co-ordinate $r$, which has the domain $[0,\infty )$.
We define a new co-ordinate $R$ by
\be
R=r^{-\frac {1}{2}},
\label{Eq-Rsoliton}
\ee
which also has the domain $[0,\infty )$.
However, $R=0$ now corresponds to infinity ($r\rightarrow \infty $),
while $R\rightarrow \infty $ corresponds to the origin ($r=0$).
This mapping of infinity to $R=0$ can be motivated from our experience
in dealing with the black hole solutions in the previous
subsection, as we already know in detail the behaviour of the
potential at $R=0$, which is crucial for applying the
multi-dimensional nodal theorem.
The specific choice of co-ordinate (\ref{Eq-Rsoliton})
will be seen below to enable us to put the potential in such
a form that the precise conditions necessary for the application
of the nodal theorem are satisfied.

With $R$ defined by (\ref{Eq-Rsoliton}),
the quantity $Q$ (\ref{Eq-Qdef}) is given by
\bdm
Q = -\frac {1}{2} r^{-\frac {3}{2}} NS .
\edm
Since this is negative, we shall use $|Q|$ in the perturbation
equations (as in (\ref{Eq-transformed1})).

As with the black hole solutions in the previous subsection,
${\cal {F}}$ (\ref{Eq-Fdef})
is regular and uniformly bounded on the whole of
$r\in [0,\infty )$ and therefore so too is the transformation
matrix $\bbR $.

For our stability proof, we require the matrix $\bbK |Q|$
to be uniformly bounded and uniformly positive definite.
However, $|Q|\sim r^{1/2}$ as $r\rightarrow \infty $,
while $\bbR $ is uniformly bounded, so
we need to introduce a second transformation matrix
$\bbY $ (see section \ref{Sect-6.4}) such that
$\bbK |Q| = \bbY \bbR ^{T} \bbR \bbY |Q|$ is
uniformly positive definite and bounded.
Let
\bdm
\bbY =\frac {1}{2{\sqrt {2}}\mu }
r^{\frac {3}{4}} \left( NS \right) ^{-1/2} \bbI ,
\edm
which satisfies the requirement that $\bbY $ should commute with
$\bbR $.
The factor $1/2{\sqrt {2}\mu }$ will be seen later in this subsection
to ensure that the behaviour of the potential as
$R\rightarrow \infty $ satisfies the conditions of the nodal theorem.
Then
\bdm
\bbK |Q| =\frac {1}{16\mu ^{2}} \bbR ^{T} \bbR
\edm
which is uniformly bounded and positive definite on
$R\in [0,\infty )$, as required.

With the additional transformation given by the matrix $\bbY $,
the potential is now ${}^{2}{\bar {\bbS }}|Q|^{-1}$, where
${}^{2}{\bar {\bbS }}$ is given by (\ref{Eq-Stilde}).
Since the matrix $\bbR $ is uniformly bounded and positive
definite, as in the previous subsection, we only need
to show that the matrix
\bdm
|Q|^{-1}\bbY
\left( \begin{array}{cc}
{}^{2}{\bar {\bbS }}_{grav} & {}^{2}{\bar {\bbS }}_{int} \\
{}^{2}{\bar {\bbS }}_{int}^{T} & {}^{2}{\bar {\bbS }}_{YM}
\end{array}
\right)\bbY
\edm
can be written in the form (\ref{Eq-nodalpotential}).
The elements of this matrix are given by (\ref{Eq-Stildebits}).
As with the black hole solutions, we consider the various
contributions to the potential in turn.

Firstly,
\bdm
\bbY = \frac {1}{2{\sqrt {2S(0)}}\mu } r^{\frac {3}{4}} \bbI
+O(r^{\frac {11}{4}})
\edm
as $r\rightarrow 0$ ($R\rightarrow \infty $)
and
\bdm
\bbY = \frac {{\sqrt {3}}}{2{\sqrt {-2\Lambda }}\mu }
r^{-\frac {1}{4}} \bbI +O(r^{-\frac {9}{4}})
\edm
as $r\rightarrow \infty $ ($R\rightarrow 0$),
where we have
used the behaviour of the metric functions in the asymptotic
regions (see section \ref{Sect-2}).
Therefore, as $r\rightarrow 0$,
\bea
\bbY \left( \bbY ^{-1} \bbY _{,\rho }  \right) \bbY |Q|^{-1}
& = & \frac {3}{16\mu ^{2} } r^{2} S(0)^{-1} \bbI + O(r^{4}) ;
\nonumber \\
\bbY  \left( \bbY ^{-1} \bbY _{,\rho \rho }  \right) \bbY |Q|^{-1}
& = &
-\frac {3}{64\mu ^{2} } r \bbI + O(r^{3}) ;
\label{Eq-Yorigin}
\eea
and as $r\rightarrow \infty $,
\bea
\bbY \left( \bbY ^{-1} \bbY _{,\rho } \right) \bbY |Q|^{-1}
& = &
\frac {3}{16\Lambda \mu ^{2} } \bbI + O(r^{-2});
\nonumber \\
\bbY  \left( \bbY ^{-1} \bbY _{,\rho \rho } \right) \bbY |Q|^{-1}
& = &
-\frac {3}{64\mu ^{2} } r \bbI + O(r^{-1}).
\nonumber
\eea
Using the fact that the matrix $\bbF $ and
its derivative with respect to $\rho $ are both regular
and uniformly bounded, for $R,r\in [0,\infty )$,
the contribution to the potential
$\bbF _{,\rho }\bbY _{2}^{-1} \bbY _{2 \, ,\rho }$
is also regular and uniformly bounded,
and so can be included in the matrix $\bbM $.
However, as $r\rightarrow \infty $ (i.e. $R\rightarrow 0$),
the matrix
$\bbY  \left( \bbY ^{-1} \bbY _{,\rho \rho }  \right) \bbY |Q|^{-1}$
diverges like $r\sim R^{-2}$.
Therefore this part of the potential will need to be included
in the $\bbL$ matrix, which we will consider in detail below.

Next we consider the contribution due to the $\bbR $
transformation, given in (\ref{Eq-Sbarbits}).
The matrix ${\bar {\bbA }}_{int}$ and
its derivative ${\bar {\bbA }}_{int \, ,\rho }$
are both regular and uniformly bounded on $r\in [0,\infty )$,
and so those terms in the potential (\ref{Eq-Sbarbits})
depending on ${\bar {\bbA }}_{int}$ can be included in the
matrix $\bbM $.

It remains to consider the following part of the potential
\bdm
|Q|^{-1} \bbY {\bar {\bbS }}\bbY = |Q|^{-1} \bbY\left( \begin{array}{cc}
{\bar {\bbS }}_{grav} & {\bar {\bbS }}_{int} \\
{\bar {\bbS }}_{int}^{T} & {\bar {\bbS }}_{YM}
\end{array}
\right)\bbY ,
\edm
whose components are given in section \ref{Sect-4}.
From our analysis of the potential for black hole solutions
(subsection \ref{Sect-6.5.1}),
we know that the potential component $S_{33}$ in
${\bar {\bbS }}_{grav}$ diverges like $r^{2}$ as $r\rightarrow \infty $,
all other terms in ${\bar {\bbS }}$ being regular.
When multiplied by $\bbY ^{2}|Q|^{-1}$, this gives a divergent
term of the form
\bdm
\frac {1}{2\mu ^{2} }r + O(r^{-1})
=\frac {1}{2\mu ^{2} } R^{-2} + O(R^{2}) .
\edm
Therefore we define a matrix $\bbL $ (\ref{Eq-nodalpotential})
to be equal to ${}^{1}\bbL +{}^{2}\bbL $, where the only non-zero entry in
${}^{1}\bbL $ is
\bdm
{}^{1}\bbL _{33} = \frac {1}{2\mu ^{2} } ;
\edm
and
\bdm
{}^{2}\bbL  = \frac {3}{64\mu ^{2} }\bbI .
\edm
Therefore the matrix $\bbL $ is uniformly bounded on
$R\in [0, \infty )$, and, furthermore, is constant.
Since $\mu ^{2} >0$, we also have that $\bbL $ is
positive definite, so in particular $\bbL (R=0)$ is positive
definite, as required by the nodal theorem.

We subtract $\bbL R^{-2}$ from the full, transformed potential
${}^{2}{\bar {\bbS }}|Q|^{-1}$ to give the matrix $\bbM $, which is then
uniformly bounded for $R\in [0,\infty )$.
The final criterion we need to check before the multi-dimensional
nodal theorem can be applied in the next subsection is that
the potential ${}^{2}{\bar {\bbS }}|Q|^{-1}$ vanishes sufficiently
quickly as $R\rightarrow \infty $ (i.e. $r\rightarrow 0$)
that
\bdm
{}^{2}{\bar {\bbS }} |Q|^{-1} R^{2} \ge -\beta _{c},
\edm
for all sufficiently large $R$,
where $\beta _{c}$ is a constant less than $1/4$.
This requires a detailed study of the behaviour of the potential
near the origin $r=0$.

We have already considered the form at the origin
of the contributions
to the potential due to the $\bbY $ transformation
(\ref{Eq-Yorigin}).
Those parts of the potential (\ref{Eq-Sbarbits})
arising from the $\bbR $ transformation, as already
observed, are bounded at the origin, and so when
multiplied by $\bbY ^{2}|Q|^{-1}\sim r^3$ as $r\rightarrow 0$,
they vanish at least as quickly as $R^{-6}$, and so need
not be considered further here.

The behaviour of the untransformed potential
${\bar {\bbS }}$ at the origin $r=0$ is
complicated due to the presence of $\gamma $ and
its derivatives.
Near the origin,
\bdm
\gamma = \frac {S(0)}{r} + O(1) .
\edm
By considering each term in the potential in turn, we
arrive at the following explicit expression for the
untransformed potential near the origin:
\bea
{\bar {\bbS }} & = &
\frac {S^{2}(0)}{r^{2}}
\left( \begin{array}{cccccc}
2-\lambda & -2{\sqrt {\lambda }} & 0 & 0 & 0 & 0
\\
-2{\sqrt {\lambda }} & \lambda & 0 & 0 & 0 & 0
\\
0 & 0 & \lambda +4 & 2 & -{\sqrt {2}}\mu  &
-{\sqrt {2\lambda }}
\\
0 & 0 & 2 & \lambda + 4 & {\sqrt {2}}\mu & -{\sqrt {2\lambda }}
\\
0 & 0 & -{\sqrt {2}}\mu & {\sqrt {2}} \mu & \lambda + 4 & 0
\\
0 & 0 & -{\sqrt {2\lambda }} & -{\sqrt {2\lambda }} & 0 & \lambda
\end{array}
\right)
+O(r^{-1})
\nonumber \\
 & = & \frac {S^{2}(0)}{r^{2}} {\cal {M}} +O(r^{-1}) ;
\nonumber
\eea
where, in the second line above,
we have defined the matrix ${\cal {M}}$.
Notice that the entry in the first row and column in the matrix ${\cal {M}}$
which is negative for $\ell\geq 2$ originates from
the $(qq)$-term in the potential ${}^{2}{\bar {\bbS }}$,
including contributions from both the untransformed
potential term $S_{22}$ (\ref{Eq-Sgrav}) and the remainder term
${\cal {S}}_{qq}$ (\ref{Eq-Sextrabits}).
Multiplying this by $\bbY ^{2} |Q|^{-1}$ and adding the contribution
to the potential from the $\bbY $ transformation (whose
behaviour near the origin is given by (\ref{Eq-Yorigin})),
we have, for the complete potential ${}^{2}{\bar {\bbS }}$,
\bdm
{}^{2} {\bar {\bbS }} = \frac {1}{4\mu ^{2} }
{\cal {M}} r + \frac {3}{64\mu ^{2} } \bbI r +O(r^{2})
=\left( \frac {1}{4\mu ^{2} }{\cal {M}} + \frac {3}{64\mu ^{2} }\bbI
\right) R^{-2} +O(R^{-4}),
\edm
as $R\rightarrow \infty $.
Therefore the potential vanishes like $R^{-2}$ as
$R\rightarrow \infty $.

In order to check whether
${}^{2}{\bar {\bbS }}|Q|^{-1}R^{2}\ge -\beta _{c}$
for all sufficiently large $R$, where $\beta _{c}<1/4$, we
need to calculate the eigenvalues of the matrix ${\cal {M}}$.
Firstly, arising from the bottom
right hand part of the matrix,
\bdm
\ell(\ell-1), \qquad
(\ell+1)(\ell+2),
\edm
both these eigenvalues repeated twice.
For $\ell\ge 1$, these eigenvalues are non-negative.
From the top left-hand part of the matrix we have
the single eigenvalues
\bdm
-\lambda , \qquad \lambda + 2 = \mu ^{2}.
\edm
One of these eigenvalues is positive, the other negative,
leading to a negative eigenvalue of ${}^{2}{\bar {\bbS }}|Q|^{-1}R^{2}$.
However, this negative eigenvalue of ${}^{2}{\bar {\bbS }}|Q|^{-1}R^{2}$
takes the form
\bdm
-\frac {1}{4} + \frac {35}{64\mu ^{2} } =-\beta _{c} >-\frac {1}{4} .
\edm
Therefore ${}^{2}{\bar {\bbS }}|Q|^{-1}R^{2}>-\beta _{c}$
for all sufficiently large $R$, where we have defined
the constant $\beta _{c}<1/4$ in the equation above.

As an aside, we note that
the eigenvector corresponding to the negative eigenvalue
of ${\cal {M}}$ has the form
\be
(q = {\sqrt {\lambda }}g , g,0,0,0,0).
\label{Eq-badevector}
\ee
We are working in a gauge in which $t=p$.
This condition does not completely specify the gauge, and
some freedom remains in our quantities $q,g,b,c,d,e$.
Physically we are interested only in gauge-invariant
quantities.
One such gauge-invariant is $\zeta $, given in (\ref{Eq-zeta}).
Since we are working in a gauge in which $t=p$, near the origin
$\zeta $ has the form
\bdm
\zeta = g-\frac {1}{{\sqrt {\lambda }}} q .
\edm
Therefore the eigenvector (\ref{Eq-badevector}) corresponds
to a perturbation in which the
gauge-invariant $\zeta $ vanishes near the origin,
and so represents a pure gauge perturbation.
The reader may wonder, at this point, whether the existence
of the negative eigenvalue in ${\cal {M}}$ is due to our
leaving part of the potential term $S_{22}$ in the remainder
${\cal {P}}$.
In fact, even if we had not performed that step, we
would still obtain a negative eigenvalue in the corresponding
${\cal {M}}$ matrix.
However, in that case the eigenvalue is considerably more
algebraically complicated than the value $-\lambda $ we
have in this case, and, furthermore, the associated
eigenvector is not so easily interpreted as being
pure gauge.

We note furthermore that, in order to construct the potential
in such a way that the conditions for the application of the
nodal theorem are satisfied, it was essential that we used
a co-ordinate $R$ which was zero at infinity ($r\rightarrow \infty $),
and tended to infinity at the origin ($r=0$).
Otherwise, we would have been forced to construct an $\bbL $
matrix such that, at $R=0$, this matrix had a negative eigenvalue,
whereas the nodal theorem requires that $\bbL $ is positive definite
at $R=0$.
By setting $R\rightarrow \infty $ at the origin $r=0$,
the matrix having the negative eigenvalue is instead the limit of
${}^{2}{\bar {\bbS }}|Q|^{-1}R^{2}$
as $R\rightarrow \infty $,
which need not be positive definite, but instead must simply be
bounded below by a constant strictly greater than $-1/4$.
Including the factor of $(2{\sqrt {2}}\mu )^{-1}$ in $\bbY $,
we have been able to ensure that the only negative
eigenvalue of this matrix at $R\rightarrow \infty $
is strictly greater than $-1/4$, thus satisfying the criteria.

To summarize, in this subsection we have shown that, for soliton
solutions, the potential has the form which enables
the application of the multi-dimensional nodal theorem.
We are now in a position to complete our stability proof.

\subsection{Analyticity argument}
\label{Sect-6.6}

In the previous subsection we cast the pulsation equations in
a form which enables us to use the multi-dimensional nodal
theorem.
From section \ref{Sect-6.3}, we know that there are
no unstable modes of the system when $|\Lambda |=\infty $.
We now wish to extend this result to sufficiently large
$|\Lambda |$.
To do this, we consider the function
${\mathfrak {F}}:=R\mapsto \det {\cal {U}}_{a_{1}}(R)$
(see section \ref{Sect-6.1} for
definitions), which we know has no nodes
in the interval $(a_{1},a_{2})$ for sufficiently
small $a_{1}$ and sufficiently large $a_{2}$ when
$|\Lambda |=\infty $.
In this subsection we shall show that ${\mathfrak {F}}$ is analytic in
$|\Lambda |$ (or, equivalently, $\xi =-\Lambda ^{-1}$),
and therefore, in a neighbourhood of $\xi =0$, the function
${\mathfrak {F}}$ will also have no zeros
(in an appropriate interval $(a_{1},a_{2})$), which means that our system
of pulsation equations has no instabilities.

Since the function ${\mathfrak {F}}$ is defined
in terms of the matrix ${\cal {U}}_{a_{1}}$, having columns
$u_{j}(R)$, where the $u_{j}$ are the solutions of the
initial value problems:
\be
{\cal {O}} u_{j} =0, \qquad a_{1}<R<\infty , \qquad u_{j}(a_{1})=0,
\qquad \frac {du_{j}}{dR}(a_{1})=e_{j},
\qquad j=1,\ldots , 6
\label{Eq-initialvalue}
\ee
(where the operator ${\cal {O}}$ is defined in (\ref{Eq-nodalform})),
${\mathfrak {F}}$ will be analytic if we can prove
that each $u_{j}(R)$ is analytic in $\xi $.
From \cite{W-Stable}, the background solutions are analytic
in $\xi $ in a neighbourhood of $\xi =0$, so all the elements
in the matrices in ${\cal {O}}$ are analytic in $\xi $ for
sufficiently small $\xi $.
The differential equation for $u_{j}$ has regular singular
points at $R=0$ and $\infty $, but all $R\in [a_{1},a_{2}]$
with $a_{1}>0$ and $a_{2}>a_{1}$
are regular points of the differential equation (\ref{Eq-initialvalue}).

Define a new independent variable ${\hat {R}}$ by
${\hat {R}}=R-a_{1}$, and make $R$ and $\xi $ into
dependent variables.
Then we consider the differential equations in (\ref{Eq-initialvalue})
(but now with ${\hat {R}}$ as the variable) together with
\bdm
\frac {dR}{d{\hat {R}}}=1, \qquad
\frac {d\xi }{d{\hat {R}}}=0.
\edm
The initial conditions are now:
\bdm
u_{j}(0)=0, \qquad
\frac {du_{j}}{d{\hat {R}}}(0)=e_{j}, \qquad
R(0)=a_{1}, \qquad
\xi (0)=\xi .
\edm
With this set-up, ${\hat {R}}=0$ is a regular point of
the system of differential equations, and so standard theorems
(see, for example, proposition 1 of \cite{breit})
tell us that we have a solution of the initial value
problem, at least in a neighbourhood of ${\hat {R}}=0$,
which is analytic in ${\hat {R}}$ and the initial parameters
$a_{1}$ and $\xi $, for all $a_{1}>0$ and $\xi $ sufficiently small.
Furthermore, since the differential equations are linear,
and the only regular singular points are at ${\hat {R}}=-a_{1}$ and
${\hat {R}}=\infty $, any solution which exists in a neighbourhood
of ${\hat {R}}=0$ can be extended to ${\hat {R}}$ sufficiently
large and positive, and will remain analytic in ${\hat {R}}$,
$a_{1}$ and $\xi $.

To summarize, for all $a_{1}>0$ and sufficiently small $\xi $,
we have solutions $u_{j}$ to the initial value problems
(\ref{Eq-initialvalue}) which are defined on $R\in [a_{1},a_{2}]$
for all $a_{2}>a_{1}$, and are analytic in $\xi $ and $R$.
Therefore, for all $a_{1}>0$ and sufficiently small $\xi $
we have a function ${\mathfrak {F}}$ which is defined on
$R\in [a_{1},a_{2}]$ for all $a_{2}>a_{1}$ and analytic in $\xi $
and $R$.
We have already shown that, when $\xi =0$, the function
${\mathfrak {F}}$ has no zeros in the interval $(a_{1},a_{2})$
for all sufficiently small $a_{1}$ and sufficiently large $a_{2}$.
We stress once again that ${\mathfrak {F}}$ must vanish
at $R=a_{1}$ (i.e. ${\hat {R}}=0$)
by the initial conditions on the $u_{j}$
(\ref{Eq-initialvalue}).
In other words, when considered as a function of ${\hat {R}}$,
the function ${\mathfrak {F}}$ has no zeros on
${\hat {R}}\in (0,\infty )$, for sufficiently small $a_{1}$, when
$\xi =0$.
We now invoke analyticity to state that for sufficiently small
$\xi $ and $a_{1}$, the function ${\mathfrak {F}}$ will
still have no zeros on ${\hat {R}}\in (0,\infty )$, i.e.
in the interval $R\in (a_{1},a_{2})$ for sufficiently large $a_{2}$.
Therefore the system of pulsation equations remains stable
for sufficiently small (non-zero) $\xi $.

\section{Conclusions}
\label{Sect-7}

This paper completes the proof that solitons and black hole
solutions to ${\mathfrak {su}}(2)$ EYM theory with a negative
cosmological constant are stable if the magnitude of the
cosmological constant is sufficiently large. Here we considered
the even-parity sector of perturbations, which proved to be
considerably more complex than the odd-parity sector studied in
\cite{SW-Stable}. It is testament to the power of the
curvature-based formalism employed here that we have been able to
complete this proof. Unlike a metric-based formulation, this
formalism yields a symmetric wave operator for the perturbations.
Using the nodal theorem, the idea is then to count the number of
nodes of the `determinant of the zero modes' which gives the
number of bound states of the spatial part of the wave operator.
Here, we have shown analytically that this determinant has no
nodes by taking the limit $\Lambda\rightarrow -\infty$. Using the
fact that the background solutions depend analytically on
$\Lambda$ we have been able to extend this result to large
$|\Lambda|$. However, by counting the nodes using a numerical
code, it would be straightforward to extend the stability analysis
to any value of $\Lambda$. Furthermore, using a numerical code, we
also expect that the non-spherical stability of the asymptotically
flat black holes and solitons \cite{bartnik,bizon} and of the
topological black holes \cite{bij} could be discussed on the same
lines as in the present article. On the other hand, our analysis
also revealed the limitations of the curvature-based formalism:
Since one does not fix the residual gauge completely, there can
still be gauge modes. For solitons, we have seen that these modes
make it difficult to write the perturbation equations in the form
required by the nodal theorem.

Physically, our result shows that the presence of a large, negative
cosmological constant stabilizes black hole hair, and also
non-Abelian solitons.
This is important in the light of the recent revival in the
study of asymptotically anti-de Sitter spacetimes,
due in part to the (rather less recent) idea of ``holography''.
The concept of holography (and its more recent relative, the
adS/CFT (conformal field theory) correspondence,
see, for example, \cite{cft} for a review)
centres around the idea that behaviour
(for example, of a supergravity theory) in the bulk
of an asymptotically adS spacetime is governed by the behaviour
of a related theory on the boundary.
In terms of these ideas, the black holes
considered in this paper are manifestly ``non-holographic'',
since their structure is not determined by quantities
(such as magnetic charge) measurable at infinity.
The consequences of the existence of these stable black holes for
the adS/CFT correspondence remain to be investigated.

On a classical and semi-classical level,
there are various open avenues of study.
Firstly, we would like to extend our current work to
seek rotating black holes and solitons in this model.
If the rotation is slow, one could use the linearized theory
to find such solutions (they are linked to the zero
modes of the system considered here). In the asymptotically flat
case, slowly rotating solutions were found in \cite{BHVS-Rotating} and some
of their rapidly rotating counterparts in \cite{KK-Rotating}.
Secondly, the behaviour of quantum fields on hairy black hole
backgrounds has received little attention to date (no doubt partly due
to the classical instability of most hairy black holes).
These questions shall be the subject of future work.

\section*{Acknowledgments}
This work was supported in part by the Swiss National Science Foundation.
E.W. would like to thank the P.P.A.R.C. (U.K.) Particle Physics Theory
Travel Fund for a grant supporting
this research, and the Center for Gravitational Physics and Geometry
at the Pennsylvania State University for hospitality.

\appendix
\section{On the stability of the RN-adS solution}
\label{App-A}

In this appendix, we discuss the positiveness of the operator
\bdm
{\cal A}_+ \equiv -\p_\rho^{\, 2} + N\left[ U + W\left(
  \begin{array}{cc} -3M & \sqrt{4G\lambda} \\ \sqrt{4G\lambda} & 3M \end{array} \right) \right]
\edm
where the functions $N$, $U$ and $W$ are given in section \ref{Sect-5.4}.

We start by showing that the function $f_{\tau}(r) = \lambda/2 + 3M/r - 2G/r^2$
which appears in the denominators of $U$ and $W$ is positive for all
$\lambda \geq 0$ and all $r \geq r_h$. Introducing the
dimensionless quantities $x = r/M$, $q^2 = G/M^2$ and $l =
-3/(\Lambda M^2)$, we can write $N$ as
\bdm
N(x) = 1 - \frac{2}{x} + \frac{q^2}{x^2} + \frac{x^2}{l}\, .
\edm
Using the fact that $N_{,x}(x_h) > 0$ and $N(x_h) = 0$, one can show
that $3 x_h - 2 q^2 > x_h^2 > 0$. Therefore, the function $3/x -
2q^2/x^2$ is positive outside the horizon, and $f_{\tau}(r) > 0$ for
all $r \geq r_h\,$.

When $\ell\geq 2$, one can generalize Chandrasekhar's arguments \cite{C-Book}
in order to relate the operator ${\cal A}_+$ to the corresponding operator
${\cal A}_-$ in the odd-parity sector by a supersymmetric-like transformation.
This works as follows:
First, we diagonalize the operator ${\cal A}_+\,$, obtaining the two
decoupled operators
\bdm
{\cal A}_{i+} = -\p_\rho^{\, 2} + N\left[ U + \sigma_i W \right]
\edm
where $\sigma_1 = -\sqrt{9M^2 + 4G\lambda}\,$, $\sigma_2 = +\sqrt{9M^2 +
4G\lambda}\,$. One can check that ${\cal A}_{i+}$ can be factorized according to
\bdm
{\cal A}_{i+} = {\cal B}_i {\cal B}_i^{\dagger} - \omega_i^2,
\edm
where
\bdm
{\cal B}_i \equiv \p_\rho + \frac{q_i N}{r(\lambda r + q_i)} + \omega_i\, ,
\;\;\;
{\cal B}_i^{\dagger} \equiv -\p_\rho + \frac{q_i N}{r(\lambda r + q_i)} + \omega_i\, ,
\edm
with $q_i = 3M - \sigma_i$ and where $\omega_i =
(\ell-1)\ell(\ell+1)(\ell+2)/2q_i$ are the algebraic special
frequencies. (Note that $(\lambda r + q_1)(\lambda r + q_2) =
2\lambda r^2 f_{\tau} > 0$, so $\lambda r + q_2$ is positive.)
The supersymmetric partners of ${\cal A}_{i+}$ are
\bdm
{\cal A}_{i-} = {\cal B}_i^{\dagger} {\cal B}_i - \omega_i^2.
\edm
Explicitly, one finds
\bdm
{\cal A}_{i-} = -\p_\rho^{\, 2} + N\left[ U_- + \sigma_i W_- \right]
\edm
where $U_- = (\ell(\ell+1) - 3M/r + 4G/r^2)/r^2$ and $W_- = 1/r^3$,
which is equivalent to the operator in the odd-parity sector!
Since we have shown in \cite{SW-Stable} that there are no unstable
modes in the odd-parity sector, and since for perturbations with
compact support, ${\cal A}_{i+}$ and ${\cal A}_{i-}$ have the same
spectrum, it follows that there are no unstable modes in the
even-parity sector either.
At this point, it is interesting to note that the functions
\bdm
u_i = \frac{e^{\omega_i\rho}}{\lambda + \frac{q_i}{r}}
\edm
fulfill ${\cal B}_i^{\dagger} u_i = 0$ and therefore are
eigenfunctions of the operators ${\cal A}_{i+}$ with negative
eigenvalue $-\omega_i^2$. However, these functions do not satisfy
homogeneous Dirichlet boundary conditions at infinity and should,
therefore, not be considered in our stability analysis.
Nevertheless, the fact that $u_i$ are finite as $r \rightarrow \infty$
could affect the equivalence of quasi-normal modes in the odd- and
even-parity sector. It has been shown recently \cite{CL1} that
-- unlike for the Schwarzschild or RN case -- the quasi-normal
frequencies of the Schwarzschild-adS solutions are different in
the odd- and even-parity sector.

Finally, we discuss the case $\ell=1$, where ${\cal A}_+$ reduces
to
\bdm
 -\p_\rho^{\, 2} + N\left[ U + 3M W \right].
\edm
Explicitly, one finds
\bdm
r^2(2f_{\tau})^2\left[ U + 3M W \right] = \frac{8}{l x^6}\left[ 9 l x^4 +
4q^4 x^4 - 18 l q^2 x^2 + 16l q^4 x - 4l q^6 \right].
\edm
Using the fact that $q^2 \leq 1$ for a black hole,
it is not difficult to show that this is positive when $x \geq 1$.
This shows the stability of all black holes with $r_h \geq M$.

\section{Factorization of the spatial operator in the odd-parity case}
\label{App-B}

In the odd-parity sector \cite{SW-Stable}, we were able to factorize
the spatial operator such that the perturbation equations assume the
form
\bdm
{\ddot{U}}_{odd} + \bbB^\dagger\bbB U_{odd} = 0,
\edm
where $U_{odd}$ is the vector containing the odd-parity perturbations.
For $\ell\geq 2$, the operator $\bbB$ is given by
\bdm
\bbB = \p_\rho + \left( \begin{array}{cc} \bbC_1 & \bbC_t^T \\
  \bbC_t & \bbD\left( \bbX_0 + T(\rho)\bbX_1 \right)\bbD - \bbA_2 \end{array} \right),
\edm
where
\bea
\bbC_1 & = & \left( \begin{array}{ccc}
  \frac{\gamma}{r} \left( \frac{r}{\gamma} \right)_{,\rho }& 0 & u \\
  0 & -\frac{\gamma_{,\rho}}{\gamma} & 0 \\
  0 & 0 & -\frac{\gamma_{,\rho}}{\gamma}
\end{array} \right),
\nonumber \\
\bbC_t^T & = & \gamma\left( \begin{array}{cccc}
  -\sqrt{\lambda} &   -v &                    0 & 0 \\
                0 & -\mu &          -\sqrt{2} w & 0 \\
                0 &    w & \frac{\mu}{\sqrt{2}} & -\sqrt{\frac{\lambda}{2}}
\end{array} \right).
\nonumber
\eea
In addition,
$\bbD = \diag(\mu\sqrt{\lambda},\mu,\sqrt{2},\mu\sqrt{\lambda/2})^{-1}$,
and the matrices $\bbX_0$, $\bbX_1$ and $\bbA_2$ are
\bea
\bbX_0 &=& \left( \begin{array}{cccc}
  -\lambda\mu^2\frac{r_{,\rho}}{r} + f_1 w\, uv & sym. & sym. & sym. \\
  -f_1 w\, u & 2w w_{,\rho }& sym. & sym. \\
   f_1 u & -2w_{,\rho }& 0 & sym. \\
   2w^2(1-w^2)u & 2w^2 w_{,\rho }& -2w w_{,\rho }& -f_2 w w_{,\rho}
\end{array} \right), \nonumber\\
\bbX_1 &=& \left( \begin{array}{cccc}
   2w^2 v^2 & sym. & sym. & sym. \\
  -2w^2 v & 2w^2 & sym. & sym. \\
   2w v & -2w & 2 & sym. \\
   f_2 w v & -f_2 w & f_2 & \frac{1}{2} f_2^2
\end{array} \right), \nonumber\\
\bbA_2 &=& -\frac{u}{\sqrt{2}}\left( \begin{array}{cccc} 0 & 0 & 0 & 1 \\ 0 & 0 & 0 & 0 \\
     0 & 0 & 0 & 0 \\ -1 & 0 & 0 & 0 \end{array} \right), \nonumber
\eea
where $f_1 = \lambda + 2w^2$ and $f_2 = \mu^2 - 2w^2$.

Finally, the function $T(\rho)$ has to satisfy the differential
equation
\be
-\p_\rho T + {\cal {A}} T^2 + {\cal {B}} T + {\cal {C}} = 0,
\label{Eq-KeyEq}
\ee
with
\bea
\mu^2\lambda {\cal {A}} &=& \frac{8G}{r^2} w^2(1-w^2)^2
+ 4\left( w^2 - 1 - \frac{\lambda}{4} \right)^2 + 4\lambda
+ \frac{7}{4}\lambda^2, \nonumber\\
\mu^2\lambda {\cal {B}} &=& 8\left[ \frac{G}{r^2}(\lambda + 2w^2)
+ 1 \right] (w^2-1)w w_{,\rho }\, ,
\nonumber \\
\mu^2\lambda {\cal {C}} &=& \left[ \frac{2G}{r^2}(\lambda + 2w^2)^2
+ 2\lambda + 4w^2 \right] w_{,\rho}^2
- \mu^2\lambda\left( \frac{\lambda}{2} + w^2 \right)\gamma^2.
\nonumber
\eea
In \cite{SW-Stable}, we have shown that equation (\ref{Eq-KeyEq})
admits global solutions, at least when $|\Lambda|$ is sufficiently
large. (There are two small errors in the expressions for ${\cal {B}}$
and ${\cal {C}}$ which are published in Ref. \cite{SW-Stable}.
However, these errors do not affect our proof that there exist
global solutions to equation (\ref{Eq-KeyEq}).)
Following the same lines, it is also easy to show that equation (\ref{Eq-KeyEq})
admits global solutions when $|w|=1$ or $w=0$.

In the sector with $\ell=1$, the matrix-valued operator $\bbB$ can even be given
explicitly, see \cite{SW-Stable}.


\begin{thebibliography}{10}

\bibitem{review}
M.S. Volkov and D.V. Gal'tsov, Phys. Reps. {\bf {319}}, 2 (1999).

\bibitem{bartnik}
R. Bartnik and J. MacKinnon, Phys. Rev. Lett. {\bf {61}}, 141
(1988).

\bibitem{bizon}
M.S. Volkov and D.V. Gal'tsov, JETP Lett. {\bf 50}, 346 (1989);
\newline
H.P. K\"unzle and A.K.M. Masood-ul-Alam, J. Math. Phys. {\bf 31},
928 (1990);
\newline
P. Bizon, Phys. Rev. Lett. {\bf 64}, 2844 (1990).

\bibitem{W-Stable}
E. Winstanley, Class. Quantum Grav. {\bf 16}, 1963 (1999).

\bibitem{bjork}
J. Bjoraker and Y. Hosotani, Phys. Rev. Lett. {\bf {84}}, 1853
(2000);
\newline
J. Bjoraker and Y. Hosotani, Phys. Rev. {\bf {D62}}, 043513
(2000).

\bibitem{brod2}
O. Brodbeck and N. Straumann, J. Math. Phys. {\bf {37}}, 1414
(1996).

\bibitem{brod1}
O. Brodbeck, M. Heusler, G. Lavrelashvili, N. Straumann and M.S.
Volkov, Phys. Rev. {\bf {D54}}, 7338 (1996).

\bibitem{SW-Stable}
O. Sarbach and E. Winstanley,
Class. Quantum Grav. {\bf 18}, 2125 (2001).

\bibitem{BHS-Letter}
O. Brodbeck, M. Heusler and O. Sarbach, Phys. Rev. Lett. {\bf 84},
3033 (2000).

\bibitem{SHB-PRD}
O. Sarbach, M. Heusler and O. Brodbeck, Phys. Rev. {\bf {D63}},
104015 (2001).

\bibitem{AQ-Nodal}
H. Amann and P. Quittner,
J. Math. Phys. {\bf 36}, 4553 (1995).

\bibitem{breit}
P. Breitenlohner, P. Forgacs and D. Maison,
Comm. Math. Phys. {\bf {163}}, 141 (1994).

\bibitem{bij}
J.J. van der Bij and E. Radu,
gr-qc/0107065.

\bibitem{Hyperbolic}
A. Anderson and J.W. York, Jr.,
Phys. Rev. Lett. {\bf 22}, 4384 (1999).

\bibitem{SHB-Odd}
O. Sarbach, M. Heusler, and O. Brodbeck,
Phys. Rev. {\bf D62}, 084001 (2000).

\bibitem{Zerilli}
F.J. Zerilli,
Phys. Rev. Lett. {\bf 24}, 737 (1970).

\bibitem{CL1}
V. Cardoso and J.P.S. Lemos,
Phys. Rev. {\bf D64}, 084017 (2001).

\bibitem{Moncrief}
V. Moncrief,
Phys. Rev. {\bf D10}, 1057 (1974);
Phys. Rev. {\bf D12}, 1526 (1975).

\bibitem{cft}
O. Aharony, S.S. Gubser, J. Maldacena, H. Ooguri and Y. Oz,
Phys. Reps. {\bf {323}}, 183 (2000).

\bibitem{BHVS-Rotating}
M.S. Volkov and N. Straumann,
Phys. Rev. Lett. {\bf 79}, 1428 (1997);
\newline
O. Brodbeck, M. Heusler, N. Straumann and M.S. Volkov,
Phys. Rev. Lett. {\bf 79}, 4310 (1997).

\bibitem{KK-Rotating}
B. Kleihaus and J. Kunz,
Phys. Rev. Lett. {\bf 86}, 3704 (2001).

\bibitem{C-Book}
S. Chandrasekhar,
{\em The Mathematical Theory of Black Holes}
(Oxford University Press, New York, 1983).


\end{thebibliography}
\end{document}